\documentclass[twocolumn, superscriptaddress, secnumarabic, amsmath, amssymb, nobibnotes, aps, prb, floatfix]{revtex4-2}
\usepackage[colorlinks=true,linkcolor=blue]{hyperref}%

\hypersetup{
  colorlinks  = true, 
  urlcolor    = blue, 
  linkcolor   = blue, 
  citecolor   = blue 
}

\usepackage{bm}
\usepackage{graphics}
\usepackage{graphicx}
\usepackage{appendix}

\newcommand{\Hami}{H}
\usepackage{xcolor}

\begin{document}

\title{Quantum Avalanches in $\mathbb{Z}_2$-preserving Interacting Ising Majorana Chain}%

\author{Lv Zhang}%
\affiliation{Institute of Physics, Chinese Academy of Sciences, Beijing 100190, China}
\affiliation{School of Physical Sciences, University of Chinese Academy of Sciences, Beijing 100049, China}

\author{Kai Xu}
\affiliation{Institute of Physics, Chinese Academy of Sciences, Beijing 100190, China}
\affiliation{Beijing Academy of Quantum Information Sciences, Beijing 100193, China}
\affiliation{Hefei National Laboratory, Hefei 230088, China}
\affiliation{Songshan Lake Materials Laboratory, Dongguan, Guangdong 523808, China}
\affiliation{CAS Center for Excellence in Topological Quantum Computation, UCAS, Beijing 100190, China}

\author{Heng Fan}
\affiliation{Institute of Physics, Chinese Academy of Sciences, Beijing 100190, China}
\affiliation{Beijing Academy of Quantum Information Sciences, Beijing 100193, China}
\affiliation{Hefei National Laboratory, Hefei 230088, China}
\affiliation{Songshan Lake Materials Laboratory, Dongguan, Guangdong 523808, China}
\affiliation{CAS Center for Excellence in Topological Quantum Computation, UCAS, Beijing 100190, China}
\date{January 10, 2025}%

\begin{abstract}
	Recent numerical works have revealed the instability of many-body localized (MBL) phase in disordered quantum many-body systems with finite system sizes and over finite timescales. 
	This instability arises from Griffith regions that occur at the thermodynamic limit, which rapidly thermalize and affect the surrounding typical MBL regions, introducing an avalanche mechanism into the system. Here, we consider the $\mathbb{Z}_2$-preserving interacting Ising Majorana chain model, which exhibits a more complex phase diagram, where an ergodic phase emerges between two MBL phases with different long-range order properties. We calculate the dynamic characteristics of the model when coupled to an infinite bath under perturbation, and through scaling behavior of the slowest thermalization rate, we find how critical disorder strengths in finite-size systems are affected by the avalanche mechanism. We also employe the embedded inclusion model and use the time evolution of mutual information between each spin and the artificial Griffith region to probe the diffusion of the thermal bubble. We observe that in finite-sized systems, the critical disorder strength gradually drifts away from the central. Our work demonstrate that both MBL paramagnetic phase and MBL spin-glass phase are unstable at finite sizes. 
\end{abstract}

\maketitle

\section{Introduction}
\label{SectionI}
Many-body localized (MBL) phase in disordered, interacting quantum many-body systems~\cite{PhysRevLett.95.206603, BASKO20061126, PhysRevB.75.155111, annurev-conmatphys-031214-014726, RevModPhys.91.021001, sierant2024manybodylocalizationageclassical} has attracted widespread attention in recent decades, as it violates the eigenstate thermalization hypothesis (ETH)~\cite{PhysRevE.50.888, doi:10.1080/00018732.2016.1198134, Rigol2008}. Anderson provided a first insight into the weak interaction limit, which can be regarded as a non-interacting problem~\cite{PhysRev.109.1492}.  Subsequent works have demonstrated that a system in MBL phase exhibits emergent quasi-integrability, and the information of the initial state is protected within the local integrals of motion (LIOM's)~\cite{PhysRevLett.111.127201, RevModPhys.91.021001, PhysRevB.90.174202, ROS2015420,PhysRevLett.117.027201, PhysRevB.97.064204, PhysRevB.106.054202} , also known as ``l-bits", and consequently the system will not reach the thermal equilibrium under its own dynamics. Recent studies have revealed that the violation of the ETH can also occur in disorder free systems, such as Stark MBL~\cite{PhysRevLett.122.040606, PhysRevLett.124.110603, Bhakuni_2020,PhysRevB.105.L140201, PhysRevB.109.024206, PhysRevB.104.205122}. Nonetheless, the later work has pointed out that the localization phenomenon in Stark MBL stems from the Hilbert space fragmentation~\cite{PhysRevB.100.214313, PhysRevX.10.011047, PhysRevX.12.011050, Moudgalya_2022}, which differs from the mechanism underlying the disordered MBL systems~\cite{PhysRevB.103.L100202}. Furthermore, a series of works have obtained the MBL phase on quantum simulation platforms~\cite{doi:10.1126/science.aaa7432, Smith2016, PhysRevLett.119.260401, PhysRevLett.120.050507, Rispoli2019, Guo2021, Leonard2023, Shi2024}.

It has now been recognized that for finite size disordered systems and over finite timescales, there may not exist a stable MBL phase, rather a MBL regime~\cite{PhysRevB.105.224203, PhysRevB.105.174205}. The results obtained from previous works, derived through exact diagonalization, do not accurately reflect the critical points at which the ergodic-MBL phase transition occurs. Although the disorder strength of a system exceeds the threshold determined by earlier researches, for sufficiently large system sizes and long time, the system may ultimately reaches its thermal equilibrium~\cite{PhysRevB.105.174205, DEROECK2023129245}. Ref.~\cite{PhysRevB.95.155129} has suggested that the instability of the MBL phase originates from local, weakly disordered regions, termed Griffith regions~\cite{Vojta2010, PhysRevB.93.134206, PhysRevLett.114.160401, andp.201600326}, which emerge in the thermodynamic limit due to statistic theory. These Griffith regions are in the thermal phase, after rapidly thermalizing within a short period of time, gradually thermalize the surrounding areas, acting as the bath, expanding the range of the thermalization regions and introducing the avalanche mechanism~\cite{PhysRevB.105.174205, PhysRevB.99.134205, PhysRevB.104.L140202,PhysRevB.106.L020202,PhysRevB.108.L020201,PhysRevResearch.5.L032011,PhysRevB.109.134202}. Recent numerical works have focused on discussing the propagation of avalanches in the typical MBL phase. One approach to probing the avalanche instability is to treat the remainder of the chain, excluding the Griffith regions, as an open system, revealing that the critical point of the disordered Heisenberg chain drifts from $W^* \approx 8$ to $W^*>20$ with increasing system sizes~\cite{PhysRevB.105.174205, PhysRevB.106.L020202}. Another approach involves modeling the bath using the same structure as the rest of the chain but with different disorder strength, placing it in the ergodic phase~\cite{PhysRevB.108.L020201, PhysRevB.109.134202}.
		
The one dimensional non-interacting Ising Majorana chain is an important model in condensed matter physics, often used to study the Majorana edge states, as an effective model of $p$-wave superconductors, introduced by Kitaev~\cite{AYuKitaev_2001}. Over the following two decades, various forms of interaction were introduced into this model~\cite{PhysRevB.81.134509, PhysRevB.83.075103, PhysRevLett.115.166401, PhysRevB.92.115137, 10.21468/SciPostPhys.14.6.152}. Recently, the $\mathbb{Z}_2$-preserving interacting Ising Majorana chain revealed a rich MBL phase diagram~\cite{PhysRevX.4.011052, PhysRevLett.113.107204, PhysRevLett.126.100604, moudgalya2020, PhysRevResearch.4.L032016, 10.21468/SciPostPhys.14.6.152}. It has been highlighted that this model has two completely different MBL phases, including trivial MBL paramagnetic (PM) phase and $\mathbb{Z}_2$ symmetry-breaking MBL spin-glass (SG) phase~\cite{moudgalya2020}. A few numerical works have shown that there is an emergent ergodic phase between two MBL phases with different long-range order properties, by calculating conventional landmarks like mean-level spacing ratio and half chain von-Neuman entropy. As the interaction strength increases, the extent of the ergodic phase continues to expand~\cite{moudgalya2020, PhysRevLett.126.100604,PhysRevResearch.4.L032016}.
	
However, it remains unclear that whether the avalanche instability affects the boundaries between each MBL phase and the ergodic phase? Will the scope of the ergodic phase expand or shrink? In this paper, to address these questions, we calculated the slowest thermalization rate of the system, coupling with an infinite bath, in the dissipationless limit, and obtained the ergodic-MBL phase transition points (for both MBL PM phase and MBL SG phase) with different system sizes~\cite{PhysRevB.105.174205, PhysRevB.106.L020202}. Meanwhile, we artificially set one end of the interacting Ising Majorana chain to be in ergodic phase~\cite{PhysRevB.108.L020201}, equivalent to a Griffith region in the thermodynamic limit, and calculated the dynamic evolution of mutual information between each spin in the rest of the chain and the effective bath. 
				
The rest of this paper is organized as follow. In Sec.~\ref{AIT}, we introduce the avalanche instability theory in quantum MBL systems and explain how Griffith regions affect the typical MBL regions. In Sec.~\ref{model}, We present the Hamiltonian of the $\mathbb{Z}_2$-preserving interacting Ising Majorana chain and discuss its phase diagram, highlighting the different eigenstate structures in the two MBL phases. In Sec.~\ref{Methods}, we provide two approaches to detect the avalanche mechanism in the Ising Majorana chain. The first method treats Griffith region as an infinite bath; the thermalization rate given by this Markov process can reflect the influence of Griffith region on a finite-length chain. The second method involves artificially placing a Griffith region at one end of the chain, and in this case, the dynamics of the typical MBL chain become a non-Markov process. We observe the diffusion of avalanches by calculating the mutual information between each spin in the chain and the effective bath. In Sec.~\ref{res}, we show the results of both Markov and non-Markov processes, which indicate that the two different MBL phases of the chain are both unstable at finite sizes. Additionally, the results from the Markov process reveal that, for small sized systems, the critical disorder of the MBL SG phase is more sensitive to the size of chain compared to that of the MBL PM phase. Finally, we summarize our findings in Sec.~\ref{sum}.
		
\section{Avalanche Instability Theory}	
\label{AIT}
In this section, we briefly introduce the avalanche instability theory~\cite{PhysRevB.95.155129, PhysRevB.105.174205} in disordered quantum MBL systems. Avalanche instability theory assumes that the MBL systems can be destabilized by a few emergent thermal inclusions, so called Griffith regions, in the thermodynamic limit. These Griffith regions thermalize rapidly, regarded as thermal bubbles, and destabilize the surrounding typical MBL chains by serving as heat reservoirs. The fate of a thermal bubble is corresponded to its thermalization rate $\Gamma$~\cite{PhysRevB.95.155129, thiery2017microscopicallymotivatedrenormalizationscheme}. Now consider a Griffith region of length $N$ within a sufficiently long one dimensional spin chain, when this thermal bubble spreads $L$ spins in each direction, the typical level spacing of this extended thermal inclusion is $2^{-(N+2L)}$ (as it has thermalized $N+2L$ spins in total). If the typical level spacing is smaller than the thermalization rate, the spin would not recognize the discreteness of the spectrum, i.e., the thermal bubble serves as a bath effectively. So the avalanche will not stop until the thermalization rate and the typical level spacing have similar energies, indicating that the scaling behavior of the thermalization rate $\Gamma\sim g^{-L}$ is important. According to the typical level spacing we have discussed before, the point where the transition occurs is at $g=4$.

The probability of obtaining a Griffith region is exponentially small in its length $N$,
\begin{eqnarray}
		P(N)\sim\mathcal{O}(e^{-N}),
\end{eqnarray}
 leading us to a reasonable assumption that there is no direct interaction between any two Griffith regions. Despite the fact that the entire long chain constitutes an isolated system, after the rapid thermalization of the Griffith region, we can focus solely on the typical MBL region and regard the Griffith region as an infinite bath which connects at one end of the MBL chain. Thus the dynamic features of the MBL chain are governed by the Master equation. The non-steady eigenmodes of the Master equation can reflect the effect of the Griffith region on MBL chain, and we can estimate the thermalization rate of the Griffith region as the slowest thermalization rate (the second largest real part eigenvalue of the Master equation). Moreover, We directly calculate the time evolution of the whole chain, referred to as the embedded thermal inclusion model~\cite{PhysRevB.108.L020201}, which involves artificially constructing a short spin chain in the thermal phase, acting as a thermal bubble connected to one end of a typical finite-size MBL chain. We probe the diffusion of the thermal bubble by studying the mutual information between each spin in typical MBL chain and the effective bath (artificial Griffith region). More details are displayed in Section.~\ref{Methods}.

\section{Model}
\label{model}
In this paper, we focus on the $\mathbb{Z}_2$-preserving interacting Ising Majorana chain model with open boundary conditions, which can be described as,
\begin{eqnarray}
	\label{HIM}
		\Hami_{\rm{IM}} = \Hami_{\rm{TFI}} + \Hami_{\rm{int}},\label{HIM}
\end{eqnarray}
where $\Hami_{\rm{TFI}}$ describes a non-interacting transverse-field Ising (TFI) chain and $\Hami_{\rm{int}}$ introduces the $\mathbb{Z}_2$ interaction in the context of Majorana fermions chain, see in Appendix~\ref{JW}, 

\begin{subequations}
	\begin{eqnarray}
		\Hami_{\rm{TFI}} = \sum_i h_i \sigma_i^z + \sum_i J_i \sigma_i^x\sigma_{i+1}^x,
		\\
		\Hami_{\rm{int}} = \sum_i V_i(\sigma_i^x\sigma_{i+2}^x + \sigma_i^z\sigma_{i+1}^z),
	\end{eqnarray}
\end{subequations}
where $ \bm{\sigma}_i = \{ \sigma_i^x,\sigma_i^y,\sigma_i^z \} $ are Pauli spin-1/2 operators at site $i=1,2,\dots,L$. $h_i,J_i$ and $V_i$ are disordered coupling strengths, they are independent random variables uniformly distributed in $[-W_h,+W_h]$, $[-W_J,+W_J]$ and $[-W_V,+W_V]$, respectively. To simplify, we focus on $W_JW_h=1$ and fix $W_V = 0.7$ in this paper.

Before studying the fate of the thermal bubbles in this model, we first review some basic properties of this interacting Ising Majorana chain. This model exhibits a $\mathbb{Z}_2$ symmetry with the parity operator $\mathbb{P}=\prod_i \sigma_i^z$, i.e., $ [\mathbb{P},\Hami_{\rm{IM}}] = 0 $. Recently, several works have revealed that this model has two distinct many body localized regions, which are separated by an emergent thermal phase, by analyzing mean-level spacing ratio~\cite{PhysRevLett.110.084101}, half chain von-Neuman entanglement entropy~\cite{LAFLORENCIE20161} and Edwards-Anderson (EA) order parameter~\cite{PhysRevB.93.134207, PhysRevLett.126.100604}. 

For $W_J \gg W_h, W_V $, the Ising interaction term dominates the Hamiltonian and the system is in the MBL SG phase, indicating that the full spectrum is pairwise degenerate~\cite{PhysRevResearch.4.L032016} as the eigenstates are so-called ``Cat state", e.g., $|\Psi\rangle = |\leftarrow\rightarrow\rightarrow\dots\rangle \pm|\rightarrow\leftarrow\leftarrow\dots\rangle$. However, for $W_h \gg W_J, W_V$, the transverse field term plays an important role, the system is in the MBL PM phase, the eigenstates are product states of the basis along the z-direction on each site, e.g., $|\Psi\rangle =|\uparrow\uparrow\downarrow\dots\rangle$.

Mean-level spacing ratio $\langle r \rangle$ illustrates two transitions around $W_J=0.32$ (from MBL PM phase to thermal phase) and $W_J=3.2$ (from thermal phase to MBL SG phase) via the finite size scaling~\cite{PhysRevLett.126.100604}. In two MBL phases, $\langle r\rangle \approx 0.39$ corresponds to the Poisson statistics~\cite{PhysRevLett.110.084101}, while in thermal phase, the energy spectrum is related to the GOE statistics with $\langle r\rangle\approx 0.53$. Unlike the mean-level spacing ratio, half chain von-Neuman entropy $S_{L/2}$ and EA order parameter can easily distinguish two MBL phases. Although $S_{L/2}$ exhibits area law both in two different MBL phases, in MBL PM phase, $S_{L/2}$ is close to $0$ and in MBL SG phase, it leads to $\ln 2 $ because the special ``Cat state" structure. In both two localized phases $S_{L/2}$ is independent of system size $L$ but in the middle thermal phase (for $W_J/W_h\approx 1$) $S_{L/2}$ increases with the system size gradually approaching the page value $S_{page}(L,L/2) = (L/2)\ln 2 -1/2$~\cite{PhysRevLett.71.1291,PhysRevLett.126.100604}. EA order parameter $\chi$ is used to distinguish the distinct MBL phase by  calculating long range Ising correlation. In MBL SG phase, $\chi$ shows the volume law as system size increase while deeply in MBL PM phase ($W_J\ll 0.32$) it approaches a constant~\cite{PhysRevLett.126.100604}.

\section{Methods}
\label{Methods}
\begin{figure}
	\includegraphics[width=1\linewidth]{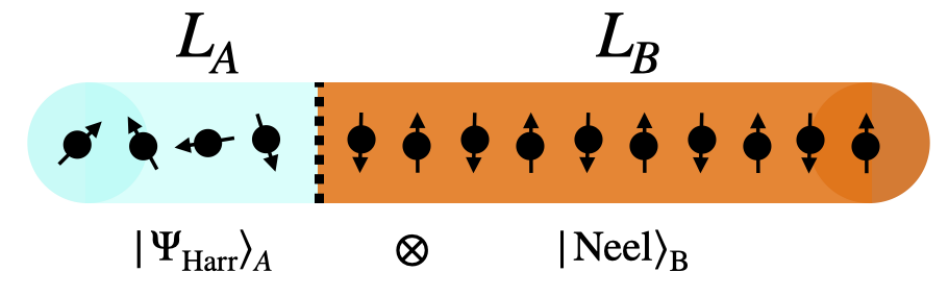}
	\caption{\label{Embedded model}Diagram depicting the setting of the embedded inclusion model. We artificially connect a Griffith region (the left part), which is in the thermal phase i.e., $W_{JA}\equiv 1$, to the one end of a typical MBL Ising Majorana chain (the right part). The initial states are prepared apart: the initial states in region A are sampled from the Harr ensemble, while in region B, the initial states are selected from two Neel states according to which MBL phase is focused. Then we time evolve the whole system by implementing the Krylov subspace algorithm. We calculate the mutual information between each spin in region B and the total region A. }
\end{figure}

\subsection{Markov Process}

To study avalanche instability as we have mentioned in Section.~\ref{SectionI}, we assume that the system couples with an infinite bath. The Master equation can then describe the dynamics of the system,
\begin{eqnarray}
	\partial_t \rho = \mathcal{L}\rho, \label{Lind}
\end{eqnarray}
where $\mathcal{L}$ is the Liouvillian superoperator,
\begin{eqnarray}
	\mathcal{L}\rho = - i[\Hami_{\rm{IM}},\rho] + \gamma \sum_{\mu} \big( L_{\mu}\rho L_{\mu}^{\dagger} -\frac{1}{2}\{ L_{\mu}^{\dagger}L_{\mu},\rho\} \big),
\end{eqnarray}
here $\{L_{\mu}\}=\{\sigma^x_1,\sigma^y_1,\sigma^z_1\}$ are Lindblad operators which describe the  couplings between the Ising Majorana chain and the infinite bath. While $\gamma$ denotes the coupling strength. We can obtain all the modes by solving for eigenoperators from,
\begin{subequations}
	\begin{eqnarray}
		\mathcal{L}\rho^R_i = \lambda_i \rho^R_i,
		\\
		\mathcal{L}^{\dagger}\rho^L_i = \lambda_i^* \rho^L_i,
	\end{eqnarray}
\end{subequations}
where $\{\rho^L_i\}$ and $\{\rho^R_i\}$ are left and right eigenmodes, $\lambda_i$ is the i$th$ eigenvalue. Thus the time dependent solution of Eq.~(\ref{Lind}) can be written as the sum of each right eigenmodes,
\begin{eqnarray}
	\rho(t) = \rho^R_0+\sum_{i>0} e^{\lambda_i t}p_i\rho_i^R,
\end{eqnarray}
where $p_i$ is the overlap between the initial state $\rho(0)=|\phi_0\rangle\langle\phi_0|$ and the i$th$ left eigenmodes i.e., $p_i= \rm{Tr}[\rho(0)\rho^L_i]$. In addition to the steady mode $\rho_0=\mathbb{I}/D$ with $\lambda_0=0$, thermalization rate of other modes is given by $\Gamma\equiv -\mathrm{Re}[\lambda]>0$. So it's natural to get slowest thermalization rate by diagonalizing the Liouvillian superoperator $\mathcal{L}$, but unfortunately, it requires a huge amount of computing resources to obtain the entire exact spectrum of a $4^L\times 4^L$ matrix, which we cannot afford even for a slightly larger system size. However, if the couplings between the infinite bath and the chain are limited (in the dissipationless limit $\gamma \ll 1$), we can reduce the problem to the subspace $\mathcal{P}$ spanned by the eigen density matrix of the original Hamiltonian Eq.~(\ref{HIM}) using perturbation theory~\cite{PhysRevB.106.L020202}. The subspace reads,
\begin{eqnarray}
	\mathcal{P} =  \bigg\{|n\rangle\langle n|\ \bigg|\  \Hami_{\rm{IM}}|n\rangle = E_n|n\rangle  \bigg\},
\end{eqnarray}
now we can estimate the slowest thermalization rate by exactly diagonalizing the reduced matrix (a $2^L\times 2^L$ non-sparse matrix) $\mathcal{S}$. The elements of the matrix $\mathcal{S}$ are 
\begin{eqnarray}
	\mathcal{S}_{nm} = \langle n |\mathcal{L}[|m\rangle\langle m|]|n\rangle =  -3\gamma\delta_{nm} + \gamma \sum_\mu |\langle n|L_{\mu}|m\rangle |^2,\label{ava_eq}
\end{eqnarray}
where $|n\rangle$ is the n$th$ eigenstate of the original interacting Ising Majorana chain Hamiltonian Eq.~(\ref{HIM}) and the slowest thermalization rate we need is the smallest non-zero eigenvalue of $\mathcal{S}$. 

Remember that the scaling behavior of the slowest thermalization rate is the vital ingredient in avalanche instability, so it's reasonable to add a scaling factor $4^L$ to the slowest thermalization rate. Therefore, what we really need is the rescaled thermalization rate $\tilde{\Gamma} = \Gamma 4^L$.

\subsection{Non-Markov Process}

We also use the embedded inclusion model~\cite{PhysRevB.108.L020201, PhysRevB.109.134202}, as shown in Fig.~\ref{Embedded model} to probe quantum avalanches, the system is divided into two parts: the left part (region A) is the interacting Ising Majorana chain in thermal phase with $W_{AJ}\equiv 1$, which behaves like a bath (Griffith region) in dynamic evolution. The right part (region B) is the Ising Majorana chain in the phase of interest, determined by varying the disorder strength $W_J$.

 To obtain how the thermal bubble spreads in the region B, we calculate the mutual information $\mathcal{I}$ at time $\tau$ between the region A (effective bath) and each spin in the region B, described as follows
 \begin{eqnarray}
 	\label{probe}
	\mathcal{I}_i(\tau) = S_{A}(\tau) + S_i(\tau) - S_{A\&i}(\tau),
\end{eqnarray}
where $S_{\Omega}= - \rm{Tr}[\rho_{\Omega}\log \rho_{\Omega}]$ is the von-Neuman entropy of the subsystem $\Omega$. Here we introduce the initial state 
\begin{eqnarray}
	|\Psi\rangle = |\Psi_{\rm{Harr}}\rangle_A \otimes |\rm{Neel}\rangle_B,
\end{eqnarray}
where $|\Psi_{\rm{Harr}}\rangle_A$ is a random state, sampled from the Harr ensemble~\cite{Roberts2017}, on region $A$. While $|\rm{Neel}\rangle_B$ denotes the $|\uparrow\downarrow\uparrow\downarrow\dots\downarrow\rangle_B$ when we focus on MBL PM phase and $|\rightarrow\leftarrow\rightarrow\leftarrow\dots\leftarrow\rangle_B$ for MBL SG phase on region B.

\section{Results}
\label{res}
\subsection{Avalanche with Markovian Dynamics}

\begin{figure}
	\includegraphics[width=1\linewidth]{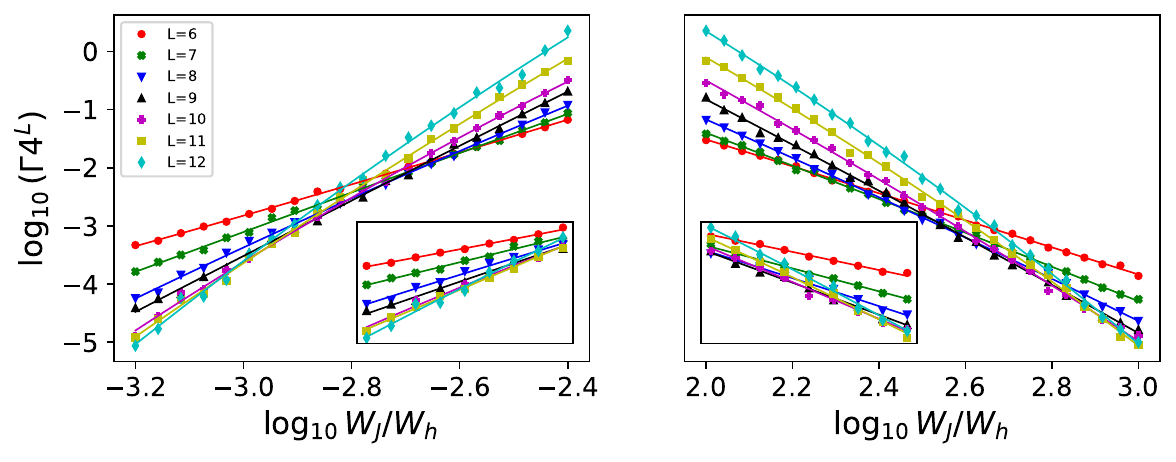}
	\caption{\label{decay_rate}Rescaled thermalization rate $\Gamma 4^L$ as a function of the disorder parameter $\lg W_J/W_h$. Points are calculated by rescaling the 80th percentile of the distribution of the smallest non-zero value of the matrix defined by Eq.~(\ref{ava_eq}) with a factor $4^L$. Curves are fitted by quadratic fitting. Different curves show different  system sizes from 6 to 12. Left panel: MBL PM phase ($\lg W_J/W_h < -2.4$), the rescaled thermalization rate increases with the disorder parameter. Right panel: MBL SG phase ($\lg W_J/W_h > 2.0$), the rescaled thermalization rate decreases with the disorder parameter. Crossing points of adjacent curves indicate the avalanche instability of different system sizes. Crossing point drifts away from the middle with the system size increasing in both two MBL phases.}
\end{figure}

We calculate the rescaled thermalization rate $\Gamma 4^L$ with multiple system sizes from 6 to 12 with at least 1200 disorder realizations, and we choose the 80th percentile of the distribution of results for each disorder parameter and system size. The results are shown in Fig.~\ref{decay_rate}.  

In Fig.~\ref{decay_rate}, it illustrates that the rescaled rate increases (decreases) with the disorder parameter $W_J/W_h$ in MBL PM phase (MBL SG phase). The solid lines in Fig.~\ref{decay_rate} are calculated by quadratic fitting for every system size with different disorder strength. Generally, we need to perform a scaling collapse of these rescaled slowest thermalization rate with different system sizes and fix disorder strength to determine the location of the phase transition. However, we don't know the exact form of the scaling function, so we use the crossing point of the two curves in Fig.~\ref{decay_rate} that the sizes of the system are adjacent to estimate the position, where the avalanche-driven transition occurs, in finite size system~\cite{PhysRevB.105.174205}. The crossing point drifts towards different directions with system size increasing, moving away from the middle, i.e., the thermalization phase diagnosed by conventional landmarks. Because of the limited acceptable system size to numerics, we can only obtain the crossing point of rescaled thermalization rate for small size system. However, we can provide an upper bound (lower bound) in MBL SG phase (MBL PM phase) in the thermodynamic limit $L\to\infty$, if we assume that the crossing point monotonically drifts with the system size $L$. In the MBL PM phase roughly at $\lg W_J/W_h < -3.05$, while in the MBL SG phase $\lg W_J/W_h > 3.07$.

\begin{figure}
	\includegraphics[width=1\linewidth]{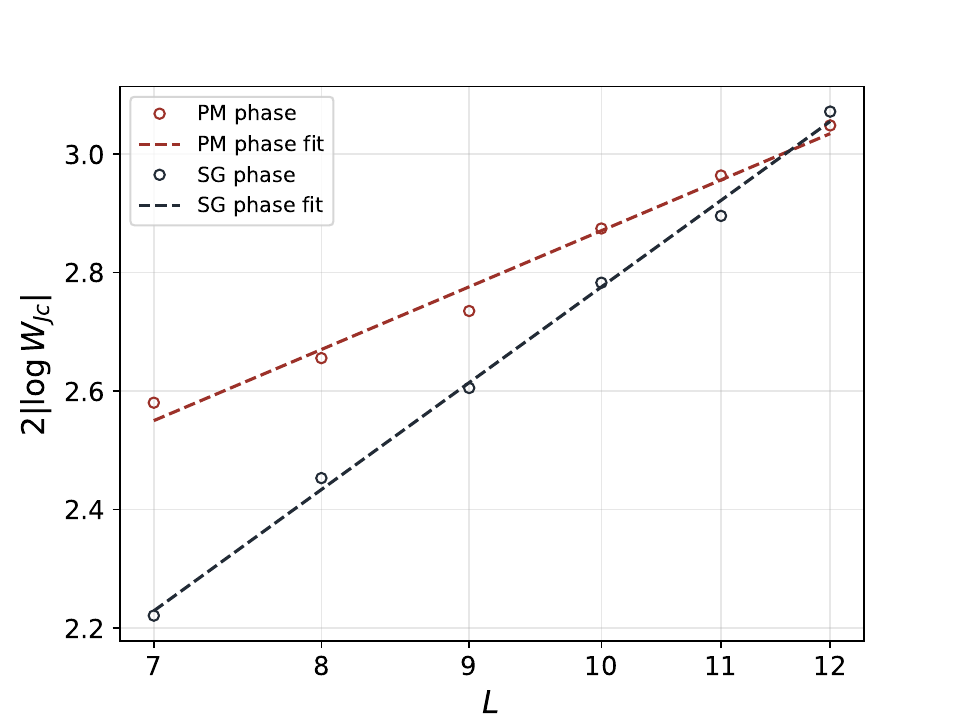}
	\caption{\label{cross} The crossing points of the rescaled thermalization rate for adjacent system sizes in Fig.~\ref{decay_rate}. The results of the MBL PM phase and MBL SG phase are displayed in one figure by taking the absolute value of $\log W_J/W_h$. The behavior of the results changs with the system size increasing, fitted by $2\zeta \log W_{JC} = \alpha \log L +\beta$ with $\zeta =1$ for MBL SG phase and $\zeta = - 1$ for MBL PM phase, showing the instability of the localization both in two MBL phases. }
\end{figure}

To show more details how the crossing point drifts with the system size, we display the crossing point $2|\lg W_{JC}|$ as a function of the system size $L$ for different MBL phases, as shown in Fig.~\ref{cross}. We use $2|\lg W_J|=\alpha \lg L + \beta$ to fit all results, which receals the slope of MBL SG phase fitting is greater than that of the MBL PM phase fitting, indicating that the critical disorder of MBL SG phase is more sensitive than that in MBL PM phase with the size of system. 

Although the $\mathbb{Z}_2$-preserving interacting Ising Majorana chain exhibits the Wannier-Kramers~\cite{PhysRev.60.252} duality transformation that allows the MBL SG phase and the MBL PM phase to correspond to each other, suggesting the emergence of a symmetric critical disorder $ W^*_{J} \Leftrightarrow -W^*_J$,  the completely different long-range order properties of its eigenstates in different MBL phases will result in distinct dynamic behaviors when the chain is connected to an infinite bath. When the chain is deeply in MBL PM phase, the eigenstates are product states of the z-direction basis at each spin, which do not exhibit long-range correlations. Therefore, the thermalization of neighboring spins by the bath in the early stage does not affect the spins at the far end. However, when the chain is in deep MBL SG phase,  the eigenstates are ``Cat-state'' along x-direction, which possess a strong long-range correlation. In this case, the thermalization of nearby spins by the bath in the early stage also affects the spins at the far end. This suggests that as we increase the size of the chain, the MBL SG phase requires a larger amount of disorder, compared to the MBL PM phase, to protect this long-range structure from being disrupted, indicating that its critical disorder is more sensitive to the size of system.

\subsection{Avalanche with Non-Markovian Dynamics}

\begin{figure*}
	\includegraphics[width=1\linewidth]{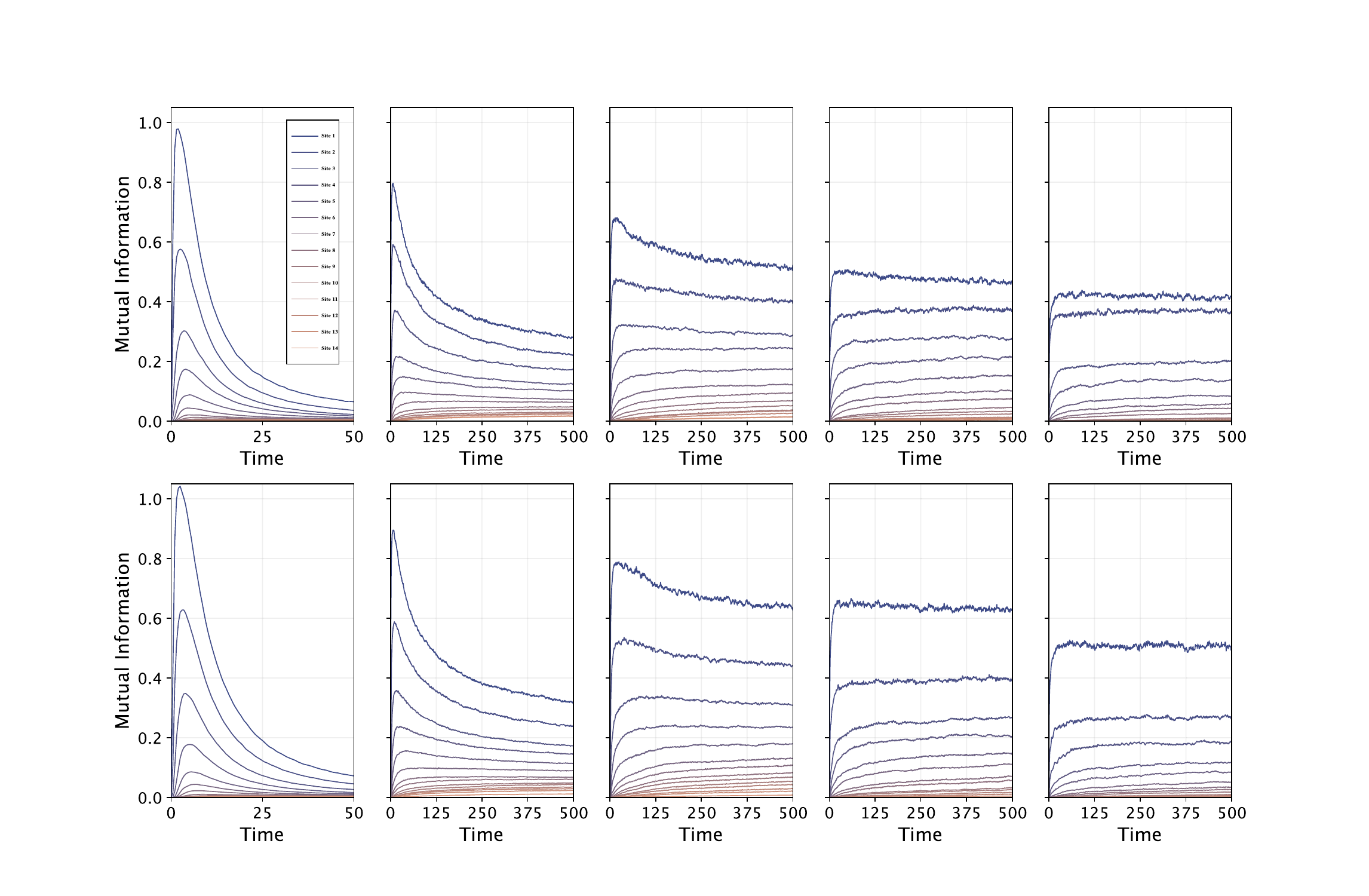}
	\caption{\label{total_dynamic}Mutual information evolution of embedded inclusion model, with region B size $L_B=14$, for several disorder strength. Top panel: results are calculated in MBL PM phase for disorder strength $W_{JB}=1,\ 0.32,\ 0.2,\ 0.128,\ 0.1$ from left to right with the initial state  $|\uparrow\downarrow\uparrow\downarrow\dots\downarrow\rangle$ in region B. Bottom panel: results are calculated in MBL SG phase for disorder strength $W_{JB}=1,\ 3.2,\ 5,\ 7.2,\ 10$ from left to right with $|\rightarrow\leftarrow\rightarrow\leftarrow\dots\leftarrow\rangle$ in region B. When $W_{JB}=1$, the whole system is in the ergodic phase, the results dsiplayed in two figures on the leftmost side illustrate the rapid thermalization in region B as the mutual information descent swiftly after it reach its maximal value in the beginning. However, as the region B gradually dives deeply into the MBL phase, the rate of decline in mutual information slows down, until the region B reaches a sufficiently high disorder strength, the mutual information would remain unchanged after a period of time evolution, indicating that the propagation of the thermal bubble fails to spread through the whole region B. The distinctly different dynamical features of mutual information show the existence of a ergodic-MBL phase transition for the system.}
\end{figure*}

In the study of the embedded inclusion model, we calculate various system sizes from 8 to 14 and 5 different disorder strengths for each MBL phase. The region B, with fixed length $L_A=4$, which is chosen to behave like a Griffith region in dynamic evolution as we set $W_{AJ}\equiv 1$. We implement the Krylov subspace algorithm~\cite{doi:10.1137/S00361445024180, PhysRevB.100.134504} to time evolve the whole model with a time step $\tau=0.5$. The results of mutual information $\mathcal{I}$ with system size $L_B=14$ are displayed in Fig.~\ref{total_dynamic}. Results of both MBL PM phase and MBL SG phase manifest the similar phenomenon with disorder strength increasing. For $W_{BJ}=1$, the whole chain is in the thermal phase, the mutual information of first several sites in the system (region B) tend to grow in the beginning, and shrink as the evolution time gets longer. However, mutual information of each site eventually converges to the same value, which is almost zero, after evolving for a sufficiently long time. We need to emphasize that this does not imply the entire system ultimately does not reach thermal equilibrium, because the mutual information between the two fixed subsystems will exponentially decrease after the fully thermalized system exceeds a certain size, see details in Appendix~\ref{appx}. Therefore, eventually converged mutual information reveals that the system and the bath are fully coupled and reach the thermalization. Meanwhile, the time when the mutual information of the first spin reaches its maximum value indicates that approximately $L_A+c$ ($c$ is a constant that is independent of $L_B$) spins are thermalized. While for region B deep in MBL phase (the rightmost column in Fig.~\ref{total_dynamic}, $W_{BJ}=10$ for MBL SG phase and $W_{BJ}=0.1$ for MBL PM phase), after a brief increase of the mutual information at the first few spins, the mutual information between each spin and the bath (region A) remains unchanged, indicating that only a limited number of spins are thermalized, i.e. the thermal bubble cannot spread to the whole system. According to the description in the introduction, we believe that this result can provide an upper bound (lower bound) for the critical point of disorder strength in MBL SG phase (MBL PM phase), specifically for the system size $L_B=14$, which influences the stability of the avalanche, $W^{\rm{crit},SG}_{BJ}\leq 10$ ($W^{\rm{crit},PM}_{BJ}\geq 0.1$).

\begin{figure*}
	\includegraphics[width=1\linewidth]{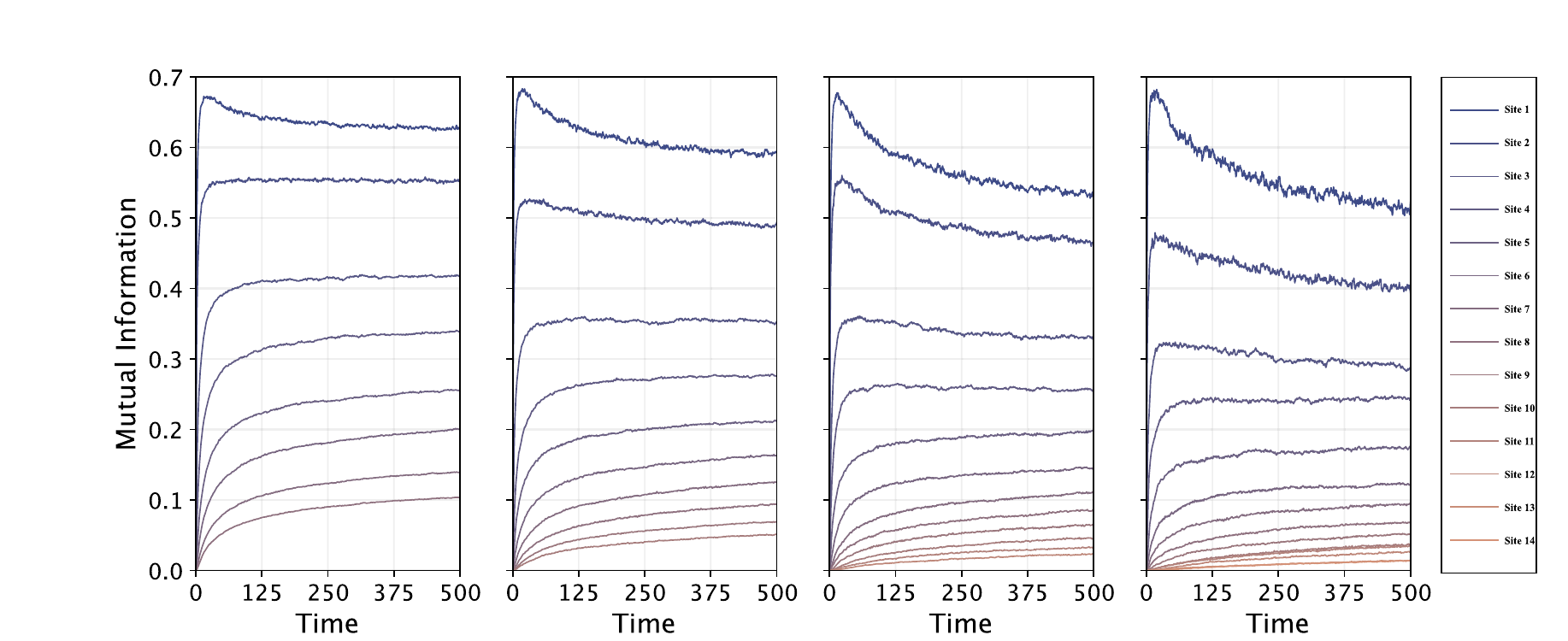}
	\caption{\label{fix_dis}Mutual information evolution of embedded inclusion model, with the same disorder strength $W_{JB}=0.2$ (deeply in the MBL PM phase according to convention landmarks), for different region B sizes. The region B sizes are chosen to be $L_B=8,10,12,14$ from left to right. For the smallest size of region B, the mutual information between the first spin and the Griffith region stay the same after a period of time, while as the size of the region B increases, the mutual information tends to descent more and more rapidly, revealing  the avalanche instability in the MBL PM phase.}
\end{figure*}

Another question we are interested in is how the size of system influences the avalanche instability, so we display different sizes of the system from 8 to 14 as shown in Fig.~\ref{fix_dis}, with the system is deep in MBL PM phase ($W_{JB}=0.2$) according to previous works. It illustrates that for the small size of the system ($L_B=8$), after $t>125$, the mutual information of each spin and the bath remains a constant, that is, the effect of the bath to the system is limited and the thermal bubble cannot spread to the entire system. However, for the lager system ($L_B=14$), the mutual information between the first spin and the bath reaches its maximum value at t=10 and then decreases rapidly, implying that the range of the avalanche of the system exceeds that of the shorter size of the system. It obviously reveals that the stability of the avalanche is not robust to the size of the system. The critical point of the disorder, which destabilizes the localization of the system, tends to drift away from the center ($W_J=1$) as system size increases.

\section{Summary}
\label{sum}
In this paper, we have explored the impacts of the avalanche instability, which arise from the Griffith regions under the thermodynamic limit, in the one dimensional $\mathbb{Z}_2$-preserving interacting Ising Majorana chain with two approaches.

We firstly treat the Griffith region as an infinite bath after a short time evolution, thus the rest of the chain is an open system and its dynamics is governed by the Master equation. By considering the dissipationless limit, we have calculated the slowest thermalization rate, which is related to the thermalization rate that stems from the Griffith regions. Our results are displayed in Fig.~\ref{decay_rate}, for each system size, the rescaled thermalization rate decreases as the system delves deeper into the MBL phase. The crossing points of the results for adjacent system sizes are the estimators of the critical points. The data is fitted using a quadratic form, and the crossing points of different MBL phases and different system sizes are shown in Fig.~\ref{cross}. The linear fitting revealed that for small system sizes, the localization of the MBL SG phase is more unstable compared to that of the MBL PM phase, as its slope is larger.

We have also directly studied the dynamic features of the original model, which is additionally connected to an artificial Griffith region. The mutual information Eq.~(\ref{probe}) between each spin out of the Griffith region and the effective bath is used to probe the entanglement and avalanches spread. The results shown in Fig.~\ref{total_dynamic} reveal that for low disorder system, thermalization gradually spreads from one end, which is connected to an artificial Griffith region, to another end, however for highly disordered (in the meaning of away from $W_J=1$) system, the erosion of thermalization stops at a particular location. These phenomena manifestly show the existence of a ergodic-MBL phase transition there. We can roughly estimate the time when the mutual information between the spin nearest the Griffith region and the effective bath begins to decline as a signature when thermalization propagates through at least $L_A+c$ spins. By displaying time evolution of mutual information for different system sizes and a fixed disorder strength, Fig.~\ref{fix_dis} illustrates that the MBL PM phase is unstable under the effects of the avalanches and as system size increases, a higher disorder is required to maintain the localized nature of the system. These conclusions are consistent with our previous analysis of the scaling behavior of the slowest thermalization rate when the system is connected to an infinite bath.     

 Our results clearly demonstrate that the critical points of ergodic-MBL (both MBL PM phase and MBL SG phase) phase transition are not stable for small system sizes and these boundaries would drift away from the central, where the system is in ergodic phase, showing the similar behavior to different models from previous works~\cite{PhysRevB.105.174205, PhysRevB.106.L020202}. Consequently, our work can deeper the understanding of the effects of avalanche in one dimensional $\mathbb{Z}_2$-preserving interacting Ising Majorana chain. Additionally, the study of the dynamic features can facilitate the  design of the experiments to directly obtain the avalanches and thermal bubbles that spread in disordered quantum many-body system.

\section*{ACKNOWLEDGMENTS}
This work was supported by the National Natural Science Foundation of China (Grants No. 92265207, No.T2121001, No.92365301, No.T2322030), the Innovation Program for Quantum Science and Technology (Grants No.2021ZD0301802), the Beijing Nova Program (No. 20220484121).

\appendix
\section{Jordan-Wigner Transformation and Ising-Majorana Chain}
\label{JW}

In this appendix, we show the details of the Majorana fermions representation. We firstly invite Jordan-Wigner transformation to convert the original spins problem into the Dirac fermions problem, the Jordan-Wigner transformation reads,
\begin{subequations}
\begin{eqnarray}
	c_i^{\dagger} &= \bigg(\prod_{k=1}^{i-1}\sigma_k^z \bigg )\sigma_i^-,
	\\
	c_i &= \bigg(\prod_{k=1}^{i-1}\sigma_k^z \bigg )\sigma_i^+,
\end{eqnarray}
\end{subequations}
where $\sigma_i^+=\frac{1}{2}(\sigma_i^x + i\sigma_i^y)$ and  $\sigma_i^-=\frac{1}{2}(\sigma_i^x - i\sigma_i^y)$, one can quickly derive $\sigma_i^z = 1 - 2c^{\dagger}_i c_i=1-2n_i$. Thus the $\mathbb{Z}_2$-preserving interacting Ising Majorana chain Hamiltonian Eq.~(\ref{HIM}) can be written as,
\begin{eqnarray}
	\begin{split}
		H_{\rm{IM}} =\ & \sum_i h_i (1 - 2n_i) 
	\\ &+ \sum_i J_i (c_i^{\dagger} - c_i) (c_{i+1}^{\dagger}+c_{i+1}) 
	\\ &+ \sum_i V_i(1-2n_i)(1-2n_{i+1})
	\\ &+ \sum_i V_i (c_i^{\dagger} - c_i)(1-2n_{i+1})(c_{i+2}^{\dagger}+c_{i+2})
	\end{split}
\end{eqnarray}
	If we rewrite each fermions as two Majorana fermions,
\begin{subequations}
\begin{eqnarray}
	\gamma_{2i-1} = c^{\dagger}_i + c_i,
	\\
	\gamma_{2i} = i(c^{\dagger}_i - c_i),
\end{eqnarray}	
\end{subequations}
then the Hamiltonian turns out to be,
\begin{subequations}
	\begin{eqnarray}    
		H_{\rm{TFI}} = -i\sum_l t_l \gamma_l \gamma_{l+1} \label{MF_TFI} 
		\\
		H_{\rm{int}} = - \sum_l g_l \gamma_l \gamma_{l+1} \gamma_{l+2} \gamma_{l+3} \label{MF_int}
	\end{eqnarray}
\end{subequations}
here $t_{2i-1}=h_i,t_{2i}=J_i$ and $g_{2i-1}=g_{2i}=V_i$. The first term Eq.~(\ref{MF_TFI}) is a standard quadratic form, thus the Hamiltonian of TFI part corresponds to a free Majorana fermions system. The second term Eq.~(\ref{MF_int}) introduces interactions between different free modes.

\section{Mutual Information in Harr Ensemble}
\label{appx}
\begin{figure}
	\includegraphics[width=1\linewidth]{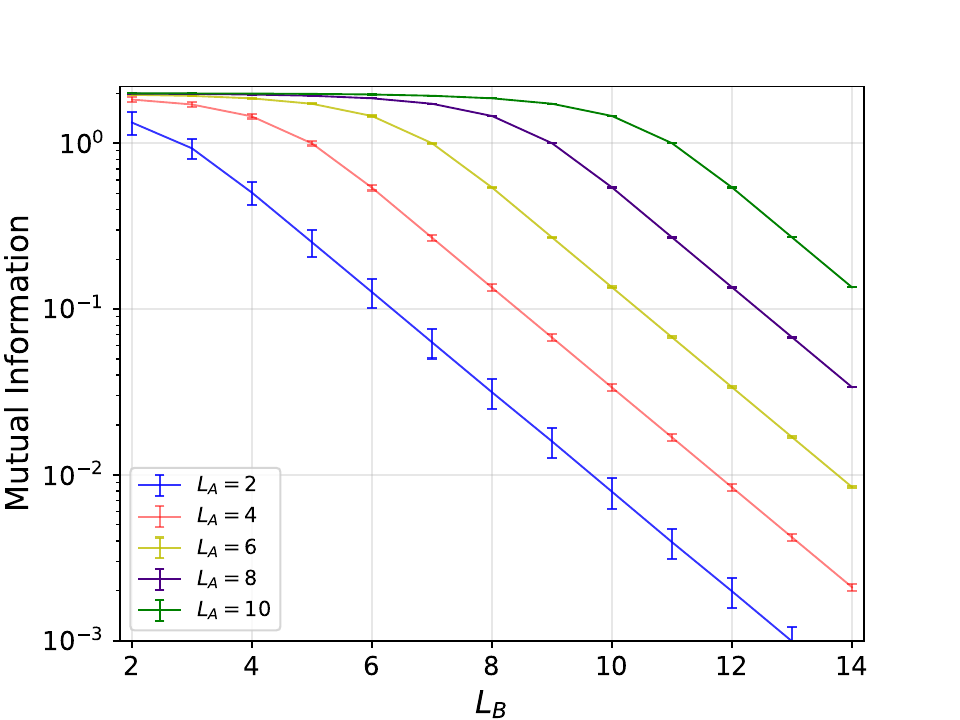}
	\caption{\label{mi} The mutual information between two subsystems of lengths $1$ and $L_A$, selected from the Harr ensemble, with the horizontal axis representing the size of the removed system (system B). After the size of system B exceeds $L_A+1$, the mutual information exhibits an exponential decay. }
\end{figure}
Consider a Harr ensemble containing $N$ spins of 1/2. If the size of a subsystem is $l$, then its Von-Neuman entanglement entropy is,
\begin{eqnarray}
	S(l|N) = n\ln 2 - \frac{1}{2^{N-2n+1}},
\end{eqnarray}
where $n=\min\{l,N-l\}$. Therefore, the mutual information between the two subsystems of sizes $L_1$ and $L_2$ in a three-part system $\Omega=A_1\otimes A_2\otimes B$ can be written as,
\begin{eqnarray}
	\mathcal{I} = S(L_1|N) + S(L_2|N) - S(L_1+L_2|N).
\end{eqnarray}

We numerically simulate the three-part system with $L_1=L_A$ and $L_2=1$, sampled from the Harr ensemble, and the results of the mutual information are displayed in Fig.~\ref{mi}. When the size of the removed subsystem $B$ is smaller than the total size of the two subsystems $A_1$ and $A_2$, the mutual information decays at a polynomial rate; however, when the size of the subsystem $B$ exceeds half of the total system size,  the mutual information exhibits an exponential decay,
\begin{eqnarray}
	\mathcal{I} \propto 2^{-L_B} \ \ \ \rm{when} \ \ \ L_B > L_A + 1.
\end{eqnarray}

Thus, when we use mutual information to probe the diffusion of the thermal bubble in the embedded inclusion model, the initial rapid increase in mutual information between the spins close to the Griffith region and the bath indicates thermal diffusion. After a sufficiently long time, the convergence of mutual information to a value close to zero shows that the system has been fully thermalized by the Griffith region, even if it is in a typical MBL phase. If the mutual information remains constant, it indicates that the system has not been fully thermalized, and the thermal bubble can only propgate a limited length.


\begin{thebibliography}{69}%
	\makeatletter
	\providecommand \@ifxundefined [1]{%
	 \@ifx{#1\undefined}
	}%
	\providecommand \@ifnum [1]{%
	 \ifnum #1\expandafter \@firstoftwo
	 \else \expandafter \@secondoftwo
	 \fi
	}%
	\providecommand \@ifx [1]{%
	 \ifx #1\expandafter \@firstoftwo
	 \else \expandafter \@secondoftwo
	 \fi
	}%
	\providecommand \natexlab [1]{#1}%
	\providecommand \enquote  [1]{``#1''}%
	\providecommand \bibnamefont  [1]{#1}%
	\providecommand \bibfnamefont [1]{#1}%
	\providecommand \citenamefont [1]{#1}%
	\providecommand \href@noop [0]{\@secondoftwo}%
	\providecommand \href [0]{\begingroup \@sanitize@url \@href}%
	\providecommand \@href[1]{\@@startlink{#1}\@@href}%
	\providecommand \@@href[1]{\endgroup#1\@@endlink}%
	\providecommand \@sanitize@url [0]{\catcode `\\12\catcode `\$12\catcode
	  `\&12\catcode `\#12\catcode `\^12\catcode `\_12\catcode `\%12\relax}%
	\providecommand \@@startlink[1]{}%
	\providecommand \@@endlink[0]{}%
	\providecommand \url  [0]{\begingroup\@sanitize@url \@url }%
	\providecommand \@url [1]{\endgroup\@href {#1}{\urlprefix }}%
	\providecommand \urlprefix  [0]{URL }%
	\providecommand \Eprint [0]{\href }%
	\providecommand \doibase [0]{https://doi.org/}%
	\providecommand \selectlanguage [0]{\@gobble}%
	\providecommand \bibinfo  [0]{\@secondoftwo}%
	\providecommand \bibfield  [0]{\@secondoftwo}%
	\providecommand \translation [1]{[#1]}%
	\providecommand \BibitemOpen [0]{}%
	\providecommand \bibitemStop [0]{}%
	\providecommand \bibitemNoStop [0]{.\EOS\space}%
	\providecommand \EOS [0]{\spacefactor3000\relax}%
	\providecommand \BibitemShut  [1]{\csname bibitem#1\endcsname}%
	\let\auto@bib@innerbib\@empty
	\bibitem [{\citenamefont {Gornyi}\ \emph {et~al.}(2005)\citenamefont {Gornyi},
	  \citenamefont {Mirlin},\ and\ \citenamefont
	  {Polyakov}}]{PhysRevLett.95.206603}%
	  \BibitemOpen
	  \bibfield  {author} {\bibinfo {author} {\bibfnamefont {I.~V.}\ \bibnamefont
	  {Gornyi}}, \bibinfo {author} {\bibfnamefont {A.~D.}\ \bibnamefont {Mirlin}},\
	  and\ \bibinfo {author} {\bibfnamefont {D.~G.}\ \bibnamefont {Polyakov}},\
	  }\bibfield  {title} {\bibinfo {title} {Interacting electrons in disordered
	  wires: Anderson localization and low-$t$ transport},\ }\href
	  {https://doi.org/10.1103/PhysRevLett.95.206603} {\bibfield  {journal}
	  {\bibinfo  {journal} {Phys. Rev. Lett.}\ }\textbf {\bibinfo {volume} {95}},\
	  \bibinfo {pages} {206603} (\bibinfo {year} {2005})}\BibitemShut {NoStop}%
	\bibitem [{\citenamefont {Basko}\ \emph {et~al.}(2006)\citenamefont {Basko},
	  \citenamefont {Aleiner},\ and\ \citenamefont {Altshuler}}]{BASKO20061126}%
	  \BibitemOpen
	  \bibfield  {author} {\bibinfo {author} {\bibfnamefont {D.}~\bibnamefont
	  {Basko}}, \bibinfo {author} {\bibfnamefont {I.}~\bibnamefont {Aleiner}},\
	  and\ \bibinfo {author} {\bibfnamefont {B.}~\bibnamefont {Altshuler}},\
	  }\bibfield  {title} {\bibinfo {title} {Metal–insulator transition in a
	  weakly interacting many-electron system with localized single-particle
	  states},\ }\href {https://doi.org/https://doi.org/10.1016/j.aop.2005.11.014}
	  {\bibfield  {journal} {\bibinfo  {journal} {Annals of Physics}\ }\textbf
	  {\bibinfo {volume} {321}},\ \bibinfo {pages} {1126} (\bibinfo {year}
	  {2006})}\BibitemShut {NoStop}%
	\bibitem [{\citenamefont {Oganesyan}\ and\ \citenamefont
	  {Huse}(2007)}]{PhysRevB.75.155111}%
	  \BibitemOpen
	  \bibfield  {author} {\bibinfo {author} {\bibfnamefont {V.}~\bibnamefont
	  {Oganesyan}}\ and\ \bibinfo {author} {\bibfnamefont {D.~A.}\ \bibnamefont
	  {Huse}},\ }\bibfield  {title} {\bibinfo {title} {Localization of interacting
	  fermions at high temperature},\ }\href
	  {https://doi.org/10.1103/PhysRevB.75.155111} {\bibfield  {journal} {\bibinfo
	  {journal} {Phys. Rev. B}\ }\textbf {\bibinfo {volume} {75}},\ \bibinfo
	  {pages} {155111} (\bibinfo {year} {2007})}\BibitemShut {NoStop}%
	\bibitem [{\citenamefont {Nandkishore}\ and\ \citenamefont
	  {Huse}(2015)}]{annurev-conmatphys-031214-014726}%
	  \BibitemOpen
	  \bibfield  {author} {\bibinfo {author} {\bibfnamefont {R.}~\bibnamefont
	  {Nandkishore}}\ and\ \bibinfo {author} {\bibfnamefont {D.~A.}\ \bibnamefont
	  {Huse}},\ }\bibfield  {title} {\bibinfo {title} {Many-body localization and
	  thermalization in quantum statistical mechanics},\ }\href
	  {https://doi.org/https://doi.org/10.1146/annurev-conmatphys-031214-014726}
	  {\bibfield  {journal} {\bibinfo  {journal} {Annual Review of Condensed Matter
	  Physics}\ }\textbf {\bibinfo {volume} {6}},\ \bibinfo {pages} {15} (\bibinfo
	  {year} {2015})}\BibitemShut {NoStop}%
	\bibitem [{\citenamefont {Abanin}\ \emph {et~al.}(2019)\citenamefont {Abanin},
	  \citenamefont {Altman}, \citenamefont {Bloch},\ and\ \citenamefont
	  {Serbyn}}]{RevModPhys.91.021001}%
	  \BibitemOpen
	  \bibfield  {author} {\bibinfo {author} {\bibfnamefont {D.~A.}\ \bibnamefont
	  {Abanin}}, \bibinfo {author} {\bibfnamefont {E.}~\bibnamefont {Altman}},
	  \bibinfo {author} {\bibfnamefont {I.}~\bibnamefont {Bloch}},\ and\ \bibinfo
	  {author} {\bibfnamefont {M.}~\bibnamefont {Serbyn}},\ }\bibfield  {title}
	  {\bibinfo {title} {Colloquium: Many-body localization, thermalization, and
	  entanglement},\ }\href {https://doi.org/10.1103/RevModPhys.91.021001}
	  {\bibfield  {journal} {\bibinfo  {journal} {Rev. Mod. Phys.}\ }\textbf
	  {\bibinfo {volume} {91}},\ \bibinfo {pages} {021001} (\bibinfo {year}
	  {2019})}\BibitemShut {NoStop}%
	\bibitem [{\citenamefont {Sierant}\ \emph {et~al.}(2024)\citenamefont
	  {Sierant}, \citenamefont {Lewenstein}, \citenamefont {Scardicchio},
	  \citenamefont {Vidmar},\ and\ \citenamefont
	  {Zakrzewski}}]{sierant2024manybodylocalizationageclassical}%
	  \BibitemOpen
	  \bibfield  {author} {\bibinfo {author} {\bibfnamefont {P.}~\bibnamefont
	  {Sierant}}, \bibinfo {author} {\bibfnamefont {M.}~\bibnamefont {Lewenstein}},
	  \bibinfo {author} {\bibfnamefont {A.}~\bibnamefont {Scardicchio}}, \bibinfo
	  {author} {\bibfnamefont {L.}~\bibnamefont {Vidmar}},\ and\ \bibinfo {author}
	  {\bibfnamefont {J.}~\bibnamefont {Zakrzewski}},\ }\href
	  {https://arxiv.org/abs/2403.07111} {\bibinfo {title} {Many-body localization
	  in the age of classical computing}} (\bibinfo {year} {2024}),\ \Eprint
	  {https://arxiv.org/abs/2403.07111} {arXiv:2403.07111 [cond-mat.dis-nn]}
	  \BibitemShut {NoStop}%
	\bibitem [{\citenamefont {Srednicki}(1994)}]{PhysRevE.50.888}%
	  \BibitemOpen
	  \bibfield  {author} {\bibinfo {author} {\bibfnamefont {M.}~\bibnamefont
	  {Srednicki}},\ }\bibfield  {title} {\bibinfo {title} {Chaos and quantum
	  thermalization},\ }\href {https://doi.org/10.1103/PhysRevE.50.888} {\bibfield
	   {journal} {\bibinfo  {journal} {Phys. Rev. E}\ }\textbf {\bibinfo {volume}
	  {50}},\ \bibinfo {pages} {888} (\bibinfo {year} {1994})}\BibitemShut
	  {NoStop}%
	\bibitem [{\citenamefont {Luca~D'Alessio}\ and\ \citenamefont
	  {Rigol}(2016)}]{doi:10.1080/00018732.2016.1198134}%
	  \BibitemOpen
	  \bibfield  {author} {\bibinfo {author} {\bibfnamefont {A.~P.}\ \bibnamefont
	  {Luca~D'Alessio}, \bibfnamefont {Yariv~Kafri}}\ and\ \bibinfo {author}
	  {\bibfnamefont {M.}~\bibnamefont {Rigol}},\ }\bibfield  {title} {\bibinfo
	  {title} {From quantum chaos and eigenstate thermalization to statistical
	  mechanics and thermodynamics},\ }\href
	  {https://doi.org/10.1080/00018732.2016.1198134} {\bibfield  {journal}
	  {\bibinfo  {journal} {Advances in Physics}\ }\textbf {\bibinfo {volume}
	  {65}},\ \bibinfo {pages} {239} (\bibinfo {year} {2016})}\BibitemShut
	  {NoStop}%
	\bibitem [{\citenamefont {Rigol}\ \emph {et~al.}(2008)\citenamefont {Rigol},
	  \citenamefont {Dunjko},\ and\ \citenamefont {Olshanii}}]{Rigol2008}%
	  \BibitemOpen
	  \bibfield  {author} {\bibinfo {author} {\bibfnamefont {M.}~\bibnamefont
	  {Rigol}}, \bibinfo {author} {\bibfnamefont {V.}~\bibnamefont {Dunjko}},\ and\
	  \bibinfo {author} {\bibfnamefont {M.}~\bibnamefont {Olshanii}},\ }\bibfield
	  {title} {\bibinfo {title} {Thermalization and its mechanism for generic
	  isolated quantum systems},\ }\href {https://doi.org/10.1038/nature06838}
	  {\bibfield  {journal} {\bibinfo  {journal} {Nature}\ }\textbf {\bibinfo
	  {volume} {452}},\ \bibinfo {pages} {854} (\bibinfo {year}
	  {2008})}\BibitemShut {NoStop}%
	\bibitem [{\citenamefont {Anderson}(1958)}]{PhysRev.109.1492}%
	  \BibitemOpen
	  \bibfield  {author} {\bibinfo {author} {\bibfnamefont {P.~W.}\ \bibnamefont
	  {Anderson}},\ }\bibfield  {title} {\bibinfo {title} {Absence of diffusion in
	  certain random lattices},\ }\href {https://doi.org/10.1103/PhysRev.109.1492}
	  {\bibfield  {journal} {\bibinfo  {journal} {Phys. Rev.}\ }\textbf {\bibinfo
	  {volume} {109}},\ \bibinfo {pages} {1492} (\bibinfo {year}
	  {1958})}\BibitemShut {NoStop}%
	\bibitem [{\citenamefont {Serbyn}\ \emph {et~al.}(2013)\citenamefont {Serbyn},
	  \citenamefont {Papi\ifmmode~\acute{c}\else \'{c}\fi{}},\ and\ \citenamefont
	  {Abanin}}]{PhysRevLett.111.127201}%
	  \BibitemOpen
	  \bibfield  {author} {\bibinfo {author} {\bibfnamefont {M.}~\bibnamefont
	  {Serbyn}}, \bibinfo {author} {\bibfnamefont {Z.}~\bibnamefont
	  {Papi\ifmmode~\acute{c}\else \'{c}\fi{}}},\ and\ \bibinfo {author}
	  {\bibfnamefont {D.~A.}\ \bibnamefont {Abanin}},\ }\bibfield  {title}
	  {\bibinfo {title} {Local conservation laws and the structure of the many-body
	  localized states},\ }\href {https://doi.org/10.1103/PhysRevLett.111.127201}
	  {\bibfield  {journal} {\bibinfo  {journal} {Phys. Rev. Lett.}\ }\textbf
	  {\bibinfo {volume} {111}},\ \bibinfo {pages} {127201} (\bibinfo {year}
	  {2013})}\BibitemShut {NoStop}%
	\bibitem [{\citenamefont {Huse}\ \emph {et~al.}(2014)\citenamefont {Huse},
	  \citenamefont {Nandkishore},\ and\ \citenamefont
	  {Oganesyan}}]{PhysRevB.90.174202}%
	  \BibitemOpen
	  \bibfield  {author} {\bibinfo {author} {\bibfnamefont {D.~A.}\ \bibnamefont
	  {Huse}}, \bibinfo {author} {\bibfnamefont {R.}~\bibnamefont {Nandkishore}},\
	  and\ \bibinfo {author} {\bibfnamefont {V.}~\bibnamefont {Oganesyan}},\
	  }\bibfield  {title} {\bibinfo {title} {Phenomenology of fully
	  many-body-localized systems},\ }\href
	  {https://doi.org/10.1103/PhysRevB.90.174202} {\bibfield  {journal} {\bibinfo
	  {journal} {Phys. Rev. B}\ }\textbf {\bibinfo {volume} {90}},\ \bibinfo
	  {pages} {174202} (\bibinfo {year} {2014})}\BibitemShut {NoStop}%
	\bibitem [{\citenamefont {Ros}\ \emph {et~al.}(2015)\citenamefont {Ros},
	  \citenamefont {Müller},\ and\ \citenamefont {Scardicchio}}]{ROS2015420}%
	  \BibitemOpen
	  \bibfield  {author} {\bibinfo {author} {\bibfnamefont {V.}~\bibnamefont
	  {Ros}}, \bibinfo {author} {\bibfnamefont {M.}~\bibnamefont {Müller}},\ and\
	  \bibinfo {author} {\bibfnamefont {A.}~\bibnamefont {Scardicchio}},\
	  }\bibfield  {title} {\bibinfo {title} {Integrals of motion in the many-body
	  localized phase},\ }\href
	  {https://doi.org/https://doi.org/10.1016/j.nuclphysb.2014.12.014} {\bibfield
	  {journal} {\bibinfo  {journal} {Nuclear Physics B}\ }\textbf {\bibinfo
	  {volume} {891}},\ \bibinfo {pages} {420} (\bibinfo {year}
	  {2015})}\BibitemShut {NoStop}%
	\bibitem [{\citenamefont {Imbrie}(2016)}]{PhysRevLett.117.027201}%
	  \BibitemOpen
	  \bibfield  {author} {\bibinfo {author} {\bibfnamefont {J.~Z.}\ \bibnamefont
	  {Imbrie}},\ }\bibfield  {title} {\bibinfo {title} {Diagonalization and
	  many-body localization for a disordered quantum spin chain},\ }\href
	  {https://doi.org/10.1103/PhysRevLett.117.027201} {\bibfield  {journal}
	  {\bibinfo  {journal} {Phys. Rev. Lett.}\ }\textbf {\bibinfo {volume} {117}},\
	  \bibinfo {pages} {027201} (\bibinfo {year} {2016})}\BibitemShut {NoStop}%
	\bibitem [{\citenamefont {Mierzejewski}\ \emph {et~al.}(2018)\citenamefont
	  {Mierzejewski}, \citenamefont {Kozarzewski},\ and\ \citenamefont
	  {Prelov\ifmmode~\check{s}\else \v{s}\fi{}ek}}]{PhysRevB.97.064204}%
	  \BibitemOpen
	  \bibfield  {author} {\bibinfo {author} {\bibfnamefont {M.}~\bibnamefont
	  {Mierzejewski}}, \bibinfo {author} {\bibfnamefont {M.}~\bibnamefont
	  {Kozarzewski}},\ and\ \bibinfo {author} {\bibfnamefont {P.}~\bibnamefont
	  {Prelov\ifmmode~\check{s}\else \v{s}\fi{}ek}},\ }\bibfield  {title} {\bibinfo
	  {title} {Counting local integrals of motion in disordered spinless-fermion
	  and hubbard chains},\ }\href {https://doi.org/10.1103/PhysRevB.97.064204}
	  {\bibfield  {journal} {\bibinfo  {journal} {Phys. Rev. B}\ }\textbf {\bibinfo
	  {volume} {97}},\ \bibinfo {pages} {064204} (\bibinfo {year}
	  {2018})}\BibitemShut {NoStop}%
	\bibitem [{\citenamefont {Adami}\ \emph {et~al.}(2022)\citenamefont {Adami},
	  \citenamefont {Amini},\ and\ \citenamefont {Soltani}}]{PhysRevB.106.054202}%
	  \BibitemOpen
	  \bibfield  {author} {\bibinfo {author} {\bibfnamefont {S.}~\bibnamefont
	  {Adami}}, \bibinfo {author} {\bibfnamefont {M.}~\bibnamefont {Amini}},\ and\
	  \bibinfo {author} {\bibfnamefont {M.}~\bibnamefont {Soltani}},\ }\bibfield
	  {title} {\bibinfo {title} {Structural properties of local integrals of motion
	  across the many-body localization transition via a fast and efficient method
	  for their construction},\ }\href
	  {https://doi.org/10.1103/PhysRevB.106.054202} {\bibfield  {journal} {\bibinfo
	   {journal} {Phys. Rev. B}\ }\textbf {\bibinfo {volume} {106}},\ \bibinfo
	  {pages} {054202} (\bibinfo {year} {2022})}\BibitemShut {NoStop}%
	\bibitem [{\citenamefont {Schulz}\ \emph {et~al.}(2019)\citenamefont {Schulz},
	  \citenamefont {Hooley}, \citenamefont {Moessner},\ and\ \citenamefont
	  {Pollmann}}]{PhysRevLett.122.040606}%
	  \BibitemOpen
	  \bibfield  {author} {\bibinfo {author} {\bibfnamefont {M.}~\bibnamefont
	  {Schulz}}, \bibinfo {author} {\bibfnamefont {C.~A.}\ \bibnamefont {Hooley}},
	  \bibinfo {author} {\bibfnamefont {R.}~\bibnamefont {Moessner}},\ and\
	  \bibinfo {author} {\bibfnamefont {F.}~\bibnamefont {Pollmann}},\ }\bibfield
	  {title} {\bibinfo {title} {Stark many-body localization},\ }\href
	  {https://doi.org/10.1103/PhysRevLett.122.040606} {\bibfield  {journal}
	  {\bibinfo  {journal} {Phys. Rev. Lett.}\ }\textbf {\bibinfo {volume} {122}},\
	  \bibinfo {pages} {040606} (\bibinfo {year} {2019})}\BibitemShut {NoStop}%
	\bibitem [{\citenamefont {Ribeiro}\ \emph {et~al.}(2020)\citenamefont
	  {Ribeiro}, \citenamefont {Lazarides},\ and\ \citenamefont
	  {Haque}}]{PhysRevLett.124.110603}%
	  \BibitemOpen
	  \bibfield  {author} {\bibinfo {author} {\bibfnamefont {P.}~\bibnamefont
	  {Ribeiro}}, \bibinfo {author} {\bibfnamefont {A.}~\bibnamefont {Lazarides}},\
	  and\ \bibinfo {author} {\bibfnamefont {M.}~\bibnamefont {Haque}},\ }\bibfield
	   {title} {\bibinfo {title} {Many-body quantum dynamics of initially trapped
	  systems due to a stark potential: Thermalization versus bloch oscillations},\
	  }\href {https://doi.org/10.1103/PhysRevLett.124.110603} {\bibfield  {journal}
	  {\bibinfo  {journal} {Phys. Rev. Lett.}\ }\textbf {\bibinfo {volume} {124}},\
	  \bibinfo {pages} {110603} (\bibinfo {year} {2020})}\BibitemShut {NoStop}%
	\bibitem [{\citenamefont {Bhakuni}\ and\ \citenamefont
	  {Sharma}(2020)}]{Bhakuni_2020}%
	  \BibitemOpen
	  \bibfield  {author} {\bibinfo {author} {\bibfnamefont {D.~S.}\ \bibnamefont
	  {Bhakuni}}\ and\ \bibinfo {author} {\bibfnamefont {A.}~\bibnamefont
	  {Sharma}},\ }\bibfield  {title} {\bibinfo {title} {Entanglement and
	  thermodynamic entropy in a clean many-body-localized system},\ }\href
	  {https://doi.org/10.1088/1361-648X/ab7c92} {\bibfield  {journal} {\bibinfo
	  {journal} {Journal of Physics: Condensed Matter}\ }\textbf {\bibinfo {volume}
	  {32}},\ \bibinfo {pages} {255603} (\bibinfo {year} {2020})}\BibitemShut
	  {NoStop}%
	\bibitem [{\citenamefont {Zisling}\ \emph {et~al.}(2022)\citenamefont
	  {Zisling}, \citenamefont {Kennes},\ and\ \citenamefont
	  {Bar~Lev}}]{PhysRevB.105.L140201}%
	  \BibitemOpen
	  \bibfield  {author} {\bibinfo {author} {\bibfnamefont {G.}~\bibnamefont
	  {Zisling}}, \bibinfo {author} {\bibfnamefont {D.~M.}\ \bibnamefont
	  {Kennes}},\ and\ \bibinfo {author} {\bibfnamefont {Y.}~\bibnamefont
	  {Bar~Lev}},\ }\bibfield  {title} {\bibinfo {title} {Transport in stark
	  many-body localized systems},\ }\href
	  {https://doi.org/10.1103/PhysRevB.105.L140201} {\bibfield  {journal}
	  {\bibinfo  {journal} {Phys. Rev. B}\ }\textbf {\bibinfo {volume} {105}},\
	  \bibinfo {pages} {L140201} (\bibinfo {year} {2022})}\BibitemShut {NoStop}%
	\bibitem [{\citenamefont {Bertoni}\ \emph {et~al.}(2024)\citenamefont
	  {Bertoni}, \citenamefont {Eisert}, \citenamefont {Kshetrimayum},
	  \citenamefont {Nietner},\ and\ \citenamefont
	  {Thomson}}]{PhysRevB.109.024206}%
	  \BibitemOpen
	  \bibfield  {author} {\bibinfo {author} {\bibfnamefont {C.}~\bibnamefont
	  {Bertoni}}, \bibinfo {author} {\bibfnamefont {J.}~\bibnamefont {Eisert}},
	  \bibinfo {author} {\bibfnamefont {A.}~\bibnamefont {Kshetrimayum}}, \bibinfo
	  {author} {\bibfnamefont {A.}~\bibnamefont {Nietner}},\ and\ \bibinfo {author}
	  {\bibfnamefont {S.~J.}\ \bibnamefont {Thomson}},\ }\bibfield  {title}
	  {\bibinfo {title} {Local integrals of motion and the stability of many-body
	  localization in wannier-stark potentials},\ }\href
	  {https://doi.org/10.1103/PhysRevB.109.024206} {\bibfield  {journal} {\bibinfo
	   {journal} {Phys. Rev. B}\ }\textbf {\bibinfo {volume} {109}},\ \bibinfo
	  {pages} {024206} (\bibinfo {year} {2024})}\BibitemShut {NoStop}%
	\bibitem [{\citenamefont {Wang}\ \emph {et~al.}(2021)\citenamefont {Wang},
	  \citenamefont {Sun},\ and\ \citenamefont {Fan}}]{PhysRevB.104.205122}%
	  \BibitemOpen
	  \bibfield  {author} {\bibinfo {author} {\bibfnamefont {Y.-Y.}\ \bibnamefont
	  {Wang}}, \bibinfo {author} {\bibfnamefont {Z.-H.}\ \bibnamefont {Sun}},\ and\
	  \bibinfo {author} {\bibfnamefont {H.}~\bibnamefont {Fan}},\ }\bibfield
	  {title} {\bibinfo {title} {Stark many-body localization transitions in
	  superconducting circuits},\ }\href
	  {https://doi.org/10.1103/PhysRevB.104.205122} {\bibfield  {journal} {\bibinfo
	   {journal} {Phys. Rev. B}\ }\textbf {\bibinfo {volume} {104}},\ \bibinfo
	  {pages} {205122} (\bibinfo {year} {2021})}\BibitemShut {NoStop}%
	\bibitem [{\citenamefont {De~Tomasi}\ \emph {et~al.}(2019)\citenamefont
	  {De~Tomasi}, \citenamefont {Hetterich}, \citenamefont {Sala},\ and\
	  \citenamefont {Pollmann}}]{PhysRevB.100.214313}%
	  \BibitemOpen
	  \bibfield  {author} {\bibinfo {author} {\bibfnamefont {G.}~\bibnamefont
	  {De~Tomasi}}, \bibinfo {author} {\bibfnamefont {D.}~\bibnamefont
	  {Hetterich}}, \bibinfo {author} {\bibfnamefont {P.}~\bibnamefont {Sala}},\
	  and\ \bibinfo {author} {\bibfnamefont {F.}~\bibnamefont {Pollmann}},\
	  }\bibfield  {title} {\bibinfo {title} {Dynamics of strongly interacting
	  systems: From fock-space fragmentation to many-body localization},\ }\href
	  {https://doi.org/10.1103/PhysRevB.100.214313} {\bibfield  {journal} {\bibinfo
	   {journal} {Phys. Rev. B}\ }\textbf {\bibinfo {volume} {100}},\ \bibinfo
	  {pages} {214313} (\bibinfo {year} {2019})}\BibitemShut {NoStop}%
	\bibitem [{\citenamefont {Sala}\ \emph {et~al.}(2020)\citenamefont {Sala},
	  \citenamefont {Rakovszky}, \citenamefont {Verresen}, \citenamefont {Knap},\
	  and\ \citenamefont {Pollmann}}]{PhysRevX.10.011047}%
	  \BibitemOpen
	  \bibfield  {author} {\bibinfo {author} {\bibfnamefont {P.}~\bibnamefont
	  {Sala}}, \bibinfo {author} {\bibfnamefont {T.}~\bibnamefont {Rakovszky}},
	  \bibinfo {author} {\bibfnamefont {R.}~\bibnamefont {Verresen}}, \bibinfo
	  {author} {\bibfnamefont {M.}~\bibnamefont {Knap}},\ and\ \bibinfo {author}
	  {\bibfnamefont {F.}~\bibnamefont {Pollmann}},\ }\bibfield  {title} {\bibinfo
	  {title} {Ergodicity breaking arising from hilbert space fragmentation in
	  dipole-conserving hamiltonians},\ }\href
	  {https://doi.org/10.1103/PhysRevX.10.011047} {\bibfield  {journal} {\bibinfo
	  {journal} {Phys. Rev. X}\ }\textbf {\bibinfo {volume} {10}},\ \bibinfo
	  {pages} {011047} (\bibinfo {year} {2020})}\BibitemShut {NoStop}%
	\bibitem [{\citenamefont {Moudgalya}\ and\ \citenamefont
	  {Motrunich}(2022)}]{PhysRevX.12.011050}%
	  \BibitemOpen
	  \bibfield  {author} {\bibinfo {author} {\bibfnamefont {S.}~\bibnamefont
	  {Moudgalya}}\ and\ \bibinfo {author} {\bibfnamefont {O.~I.}\ \bibnamefont
	  {Motrunich}},\ }\bibfield  {title} {\bibinfo {title} {Hilbert space
	  fragmentation and commutant algebras},\ }\href
	  {https://doi.org/10.1103/PhysRevX.12.011050} {\bibfield  {journal} {\bibinfo
	  {journal} {Phys. Rev. X}\ }\textbf {\bibinfo {volume} {12}},\ \bibinfo
	  {pages} {011050} (\bibinfo {year} {2022})}\BibitemShut {NoStop}%
	\bibitem [{\citenamefont {Moudgalya}\ \emph {et~al.}(2022)\citenamefont
	  {Moudgalya}, \citenamefont {Bernevig},\ and\ \citenamefont
	  {Regnault}}]{Moudgalya_2022}%
	  \BibitemOpen
	  \bibfield  {author} {\bibinfo {author} {\bibfnamefont {S.}~\bibnamefont
	  {Moudgalya}}, \bibinfo {author} {\bibfnamefont {B.~A.}\ \bibnamefont
	  {Bernevig}},\ and\ \bibinfo {author} {\bibfnamefont {N.}~\bibnamefont
	  {Regnault}},\ }\bibfield  {title} {\bibinfo {title} {Quantum many-body scars
	  and hilbert space fragmentation: a review of exact results},\ }\href
	  {https://doi.org/10.1088/1361-6633/ac73a0} {\bibfield  {journal} {\bibinfo
	  {journal} {Reports on Progress in Physics}\ }\textbf {\bibinfo {volume}
	  {85}},\ \bibinfo {pages} {086501} (\bibinfo {year} {2022})}\BibitemShut
	  {NoStop}%
	\bibitem [{\citenamefont {Doggen}\ \emph {et~al.}(2021)\citenamefont {Doggen},
	  \citenamefont {Gornyi},\ and\ \citenamefont
	  {Polyakov}}]{PhysRevB.103.L100202}%
	  \BibitemOpen
	  \bibfield  {author} {\bibinfo {author} {\bibfnamefont {E.~V.~H.}\
	  \bibnamefont {Doggen}}, \bibinfo {author} {\bibfnamefont {I.~V.}\
	  \bibnamefont {Gornyi}},\ and\ \bibinfo {author} {\bibfnamefont {D.~G.}\
	  \bibnamefont {Polyakov}},\ }\bibfield  {title} {\bibinfo {title} {Stark
	  many-body localization: Evidence for hilbert-space shattering},\ }\href
	  {https://doi.org/10.1103/PhysRevB.103.L100202} {\bibfield  {journal}
	  {\bibinfo  {journal} {Phys. Rev. B}\ }\textbf {\bibinfo {volume} {103}},\
	  \bibinfo {pages} {L100202} (\bibinfo {year} {2021})}\BibitemShut {NoStop}%
	\bibitem [{\citenamefont {Schreiber}\ \emph {et~al.}(2015)\citenamefont
	  {Schreiber}, \citenamefont {Hodgman}, \citenamefont {Bordia}, \citenamefont
	  {Lüschen}, \citenamefont {Fischer}, \citenamefont {Vosk}, \citenamefont
	  {Altman}, \citenamefont {Schneider},\ and\ \citenamefont
	  {Bloch}}]{doi:10.1126/science.aaa7432}%
	  \BibitemOpen
	  \bibfield  {author} {\bibinfo {author} {\bibfnamefont {M.}~\bibnamefont
	  {Schreiber}}, \bibinfo {author} {\bibfnamefont {S.~S.}\ \bibnamefont
	  {Hodgman}}, \bibinfo {author} {\bibfnamefont {P.}~\bibnamefont {Bordia}},
	  \bibinfo {author} {\bibfnamefont {H.~P.}\ \bibnamefont {Lüschen}}, \bibinfo
	  {author} {\bibfnamefont {M.~H.}\ \bibnamefont {Fischer}}, \bibinfo {author}
	  {\bibfnamefont {R.}~\bibnamefont {Vosk}}, \bibinfo {author} {\bibfnamefont
	  {E.}~\bibnamefont {Altman}}, \bibinfo {author} {\bibfnamefont
	  {U.}~\bibnamefont {Schneider}},\ and\ \bibinfo {author} {\bibfnamefont
	  {I.}~\bibnamefont {Bloch}},\ }\bibfield  {title} {\bibinfo {title}
	  {Observation of many-body localization of interacting fermions in a
	  quasirandom optical lattice},\ }\href
	  {https://doi.org/10.1126/science.aaa7432} {\bibfield  {journal} {\bibinfo
	  {journal} {Science}\ }\textbf {\bibinfo {volume} {349}},\ \bibinfo {pages}
	  {842} (\bibinfo {year} {2015})}\BibitemShut {NoStop}%
	\bibitem [{\citenamefont {Smith}\ \emph {et~al.}(2016)\citenamefont {Smith},
	  \citenamefont {Lee}, \citenamefont {Richerme}, \citenamefont {Neyenhuis},
	  \citenamefont {Hess}, \citenamefont {Hauke}, \citenamefont {Heyl},
	  \citenamefont {Huse},\ and\ \citenamefont {Monroe}}]{Smith2016}%
	  \BibitemOpen
	  \bibfield  {author} {\bibinfo {author} {\bibfnamefont {J.}~\bibnamefont
	  {Smith}}, \bibinfo {author} {\bibfnamefont {A.}~\bibnamefont {Lee}}, \bibinfo
	  {author} {\bibfnamefont {P.}~\bibnamefont {Richerme}}, \bibinfo {author}
	  {\bibfnamefont {B.}~\bibnamefont {Neyenhuis}}, \bibinfo {author}
	  {\bibfnamefont {P.~W.}\ \bibnamefont {Hess}}, \bibinfo {author}
	  {\bibfnamefont {P.}~\bibnamefont {Hauke}}, \bibinfo {author} {\bibfnamefont
	  {M.}~\bibnamefont {Heyl}}, \bibinfo {author} {\bibfnamefont {D.~A.}\
	  \bibnamefont {Huse}},\ and\ \bibinfo {author} {\bibfnamefont
	  {C.}~\bibnamefont {Monroe}},\ }\bibfield  {title} {\bibinfo {title}
	  {Many-body localization in a quantum simulator with programmable random
	  disorder},\ }\href {https://doi.org/10.1038/nphys3783} {\bibfield  {journal}
	  {\bibinfo  {journal} {Nature Physics}\ }\textbf {\bibinfo {volume} {12}},\
	  \bibinfo {pages} {907} (\bibinfo {year} {2016})}\BibitemShut {NoStop}%
	\bibitem [{\citenamefont {L\"uschen}\ \emph {et~al.}(2017)\citenamefont
	  {L\"uschen}, \citenamefont {Bordia}, \citenamefont {Scherg}, \citenamefont
	  {Alet}, \citenamefont {Altman}, \citenamefont {Schneider},\ and\
	  \citenamefont {Bloch}}]{PhysRevLett.119.260401}%
	  \BibitemOpen
	  \bibfield  {author} {\bibinfo {author} {\bibfnamefont {H.~P.}\ \bibnamefont
	  {L\"uschen}}, \bibinfo {author} {\bibfnamefont {P.}~\bibnamefont {Bordia}},
	  \bibinfo {author} {\bibfnamefont {S.}~\bibnamefont {Scherg}}, \bibinfo
	  {author} {\bibfnamefont {F.}~\bibnamefont {Alet}}, \bibinfo {author}
	  {\bibfnamefont {E.}~\bibnamefont {Altman}}, \bibinfo {author} {\bibfnamefont
	  {U.}~\bibnamefont {Schneider}},\ and\ \bibinfo {author} {\bibfnamefont
	  {I.}~\bibnamefont {Bloch}},\ }\bibfield  {title} {\bibinfo {title}
	  {Observation of slow dynamics near the many-body localization transition in
	  one-dimensional quasiperiodic systems},\ }\href
	  {https://doi.org/10.1103/PhysRevLett.119.260401} {\bibfield  {journal}
	  {\bibinfo  {journal} {Phys. Rev. Lett.}\ }\textbf {\bibinfo {volume} {119}},\
	  \bibinfo {pages} {260401} (\bibinfo {year} {2017})}\BibitemShut {NoStop}%
	\bibitem [{\citenamefont {Xu}\ \emph {et~al.}(2018)\citenamefont {Xu},
	  \citenamefont {Chen}, \citenamefont {Zeng}, \citenamefont {Zhang},
	  \citenamefont {Song}, \citenamefont {Liu}, \citenamefont {Guo}, \citenamefont
	  {Zhang}, \citenamefont {Xu}, \citenamefont {Deng}, \citenamefont {Huang},
	  \citenamefont {Wang}, \citenamefont {Zhu}, \citenamefont {Zheng},\ and\
	  \citenamefont {Fan}}]{PhysRevLett.120.050507}%
	  \BibitemOpen
	  \bibfield  {author} {\bibinfo {author} {\bibfnamefont {K.}~\bibnamefont
	  {Xu}}, \bibinfo {author} {\bibfnamefont {J.-J.}\ \bibnamefont {Chen}},
	  \bibinfo {author} {\bibfnamefont {Y.}~\bibnamefont {Zeng}}, \bibinfo {author}
	  {\bibfnamefont {Y.-R.}\ \bibnamefont {Zhang}}, \bibinfo {author}
	  {\bibfnamefont {C.}~\bibnamefont {Song}}, \bibinfo {author} {\bibfnamefont
	  {W.}~\bibnamefont {Liu}}, \bibinfo {author} {\bibfnamefont {Q.}~\bibnamefont
	  {Guo}}, \bibinfo {author} {\bibfnamefont {P.}~\bibnamefont {Zhang}}, \bibinfo
	  {author} {\bibfnamefont {D.}~\bibnamefont {Xu}}, \bibinfo {author}
	  {\bibfnamefont {H.}~\bibnamefont {Deng}}, \bibinfo {author} {\bibfnamefont
	  {K.}~\bibnamefont {Huang}}, \bibinfo {author} {\bibfnamefont
	  {H.}~\bibnamefont {Wang}}, \bibinfo {author} {\bibfnamefont {X.}~\bibnamefont
	  {Zhu}}, \bibinfo {author} {\bibfnamefont {D.}~\bibnamefont {Zheng}},\ and\
	  \bibinfo {author} {\bibfnamefont {H.}~\bibnamefont {Fan}},\ }\bibfield
	  {title} {\bibinfo {title} {Emulating many-body localization with a
	  superconducting quantum processor},\ }\href
	  {https://doi.org/10.1103/PhysRevLett.120.050507} {\bibfield  {journal}
	  {\bibinfo  {journal} {Phys. Rev. Lett.}\ }\textbf {\bibinfo {volume} {120}},\
	  \bibinfo {pages} {050507} (\bibinfo {year} {2018})}\BibitemShut {NoStop}%
	\bibitem [{\citenamefont {Rispoli}\ \emph {et~al.}(2019)\citenamefont
	  {Rispoli}, \citenamefont {Lukin}, \citenamefont {Schittko}, \citenamefont
	  {Kim}, \citenamefont {Tai}, \citenamefont {L{\'e}onard},\ and\ \citenamefont
	  {Greiner}}]{Rispoli2019}%
	  \BibitemOpen
	  \bibfield  {author} {\bibinfo {author} {\bibfnamefont {M.}~\bibnamefont
	  {Rispoli}}, \bibinfo {author} {\bibfnamefont {A.}~\bibnamefont {Lukin}},
	  \bibinfo {author} {\bibfnamefont {R.}~\bibnamefont {Schittko}}, \bibinfo
	  {author} {\bibfnamefont {S.}~\bibnamefont {Kim}}, \bibinfo {author}
	  {\bibfnamefont {M.~E.}\ \bibnamefont {Tai}}, \bibinfo {author} {\bibfnamefont
	  {J.}~\bibnamefont {L{\'e}onard}},\ and\ \bibinfo {author} {\bibfnamefont
	  {M.}~\bibnamefont {Greiner}},\ }\bibfield  {title} {\bibinfo {title} {Quantum
	  critical behaviour at the many-body localization transition},\ }\href
	  {https://doi.org/10.1038/s41586-019-1527-2} {\bibfield  {journal} {\bibinfo
	  {journal} {Nature}\ }\textbf {\bibinfo {volume} {573}},\ \bibinfo {pages}
	  {385} (\bibinfo {year} {2019})}\BibitemShut {NoStop}%
	\bibitem [{\citenamefont {Guo}\ \emph {et~al.}(2021)\citenamefont {Guo},
	  \citenamefont {Cheng}, \citenamefont {Sun}, \citenamefont {Song},
	  \citenamefont {Li}, \citenamefont {Wang}, \citenamefont {Ren}, \citenamefont
	  {Dong}, \citenamefont {Zheng}, \citenamefont {Zhang}, \citenamefont
	  {Mondaini}, \citenamefont {Fan},\ and\ \citenamefont {Wang}}]{Guo2021}%
	  \BibitemOpen
	  \bibfield  {author} {\bibinfo {author} {\bibfnamefont {Q.}~\bibnamefont
	  {Guo}}, \bibinfo {author} {\bibfnamefont {C.}~\bibnamefont {Cheng}}, \bibinfo
	  {author} {\bibfnamefont {Z.-H.}\ \bibnamefont {Sun}}, \bibinfo {author}
	  {\bibfnamefont {Z.}~\bibnamefont {Song}}, \bibinfo {author} {\bibfnamefont
	  {H.}~\bibnamefont {Li}}, \bibinfo {author} {\bibfnamefont {Z.}~\bibnamefont
	  {Wang}}, \bibinfo {author} {\bibfnamefont {W.}~\bibnamefont {Ren}}, \bibinfo
	  {author} {\bibfnamefont {H.}~\bibnamefont {Dong}}, \bibinfo {author}
	  {\bibfnamefont {D.}~\bibnamefont {Zheng}}, \bibinfo {author} {\bibfnamefont
	  {Y.-R.}\ \bibnamefont {Zhang}}, \bibinfo {author} {\bibfnamefont
	  {R.}~\bibnamefont {Mondaini}}, \bibinfo {author} {\bibfnamefont
	  {H.}~\bibnamefont {Fan}},\ and\ \bibinfo {author} {\bibfnamefont
	  {H.}~\bibnamefont {Wang}},\ }\bibfield  {title} {\bibinfo {title}
	  {Observation of energy-resolved many-body localization},\ }\href
	  {https://doi.org/10.1038/s41567-020-1035-1} {\bibfield  {journal} {\bibinfo
	  {journal} {Nature Physics}\ }\textbf {\bibinfo {volume} {17}},\ \bibinfo
	  {pages} {234} (\bibinfo {year} {2021})}\BibitemShut {NoStop}%
	\bibitem [{\citenamefont {L{\'e}onard}\ \emph {et~al.}(2023)\citenamefont
	  {L{\'e}onard}, \citenamefont {Kim}, \citenamefont {Rispoli}, \citenamefont
	  {Lukin}, \citenamefont {Schittko}, \citenamefont {Kwan}, \citenamefont
	  {Demler}, \citenamefont {Sels},\ and\ \citenamefont {Greiner}}]{Leonard2023}%
	  \BibitemOpen
	  \bibfield  {author} {\bibinfo {author} {\bibfnamefont {J.}~\bibnamefont
	  {L{\'e}onard}}, \bibinfo {author} {\bibfnamefont {S.}~\bibnamefont {Kim}},
	  \bibinfo {author} {\bibfnamefont {M.}~\bibnamefont {Rispoli}}, \bibinfo
	  {author} {\bibfnamefont {A.}~\bibnamefont {Lukin}}, \bibinfo {author}
	  {\bibfnamefont {R.}~\bibnamefont {Schittko}}, \bibinfo {author}
	  {\bibfnamefont {J.}~\bibnamefont {Kwan}}, \bibinfo {author} {\bibfnamefont
	  {E.}~\bibnamefont {Demler}}, \bibinfo {author} {\bibfnamefont
	  {D.}~\bibnamefont {Sels}},\ and\ \bibinfo {author} {\bibfnamefont
	  {M.}~\bibnamefont {Greiner}},\ }\bibfield  {title} {\bibinfo {title} {Probing
	  the onset of quantum avalanches in a many-body localized system},\ }\href
	  {https://doi.org/10.1038/s41567-022-01887-3} {\bibfield  {journal} {\bibinfo
	  {journal} {Nature Physics}\ }\textbf {\bibinfo {volume} {19}},\ \bibinfo
	  {pages} {481} (\bibinfo {year} {2023})}\BibitemShut {NoStop}%
	\bibitem [{\citenamefont {Shi}\ \emph {et~al.}(2024)\citenamefont {Shi},
	  \citenamefont {Sun}, \citenamefont {Wang}, \citenamefont {Wang},
	  \citenamefont {Zhang}, \citenamefont {Ma}, \citenamefont {Liu}, \citenamefont
	  {Zhao}, \citenamefont {Song}, \citenamefont {Liang}, \citenamefont {Mei},
	  \citenamefont {Zhang}, \citenamefont {Li}, \citenamefont {Chen},
	  \citenamefont {Song}, \citenamefont {Wang}, \citenamefont {Xue},
	  \citenamefont {Yu}, \citenamefont {Huang}, \citenamefont {Xiang},
	  \citenamefont {Xu}, \citenamefont {Zheng},\ and\ \citenamefont
	  {Fan}}]{Shi2024}%
	  \BibitemOpen
	  \bibfield  {author} {\bibinfo {author} {\bibfnamefont {Y.-H.}\ \bibnamefont
	  {Shi}}, \bibinfo {author} {\bibfnamefont {Z.-H.}\ \bibnamefont {Sun}},
	  \bibinfo {author} {\bibfnamefont {Y.-Y.}\ \bibnamefont {Wang}}, \bibinfo
	  {author} {\bibfnamefont {Z.-A.}\ \bibnamefont {Wang}}, \bibinfo {author}
	  {\bibfnamefont {Y.-R.}\ \bibnamefont {Zhang}}, \bibinfo {author}
	  {\bibfnamefont {W.-G.}\ \bibnamefont {Ma}}, \bibinfo {author} {\bibfnamefont
	  {H.-T.}\ \bibnamefont {Liu}}, \bibinfo {author} {\bibfnamefont
	  {K.}~\bibnamefont {Zhao}}, \bibinfo {author} {\bibfnamefont {J.-C.}\
	  \bibnamefont {Song}}, \bibinfo {author} {\bibfnamefont {G.-H.}\ \bibnamefont
	  {Liang}}, \bibinfo {author} {\bibfnamefont {Z.-Y.}\ \bibnamefont {Mei}},
	  \bibinfo {author} {\bibfnamefont {J.-C.}\ \bibnamefont {Zhang}}, \bibinfo
	  {author} {\bibfnamefont {H.}~\bibnamefont {Li}}, \bibinfo {author}
	  {\bibfnamefont {C.-T.}\ \bibnamefont {Chen}}, \bibinfo {author}
	  {\bibfnamefont {X.}~\bibnamefont {Song}}, \bibinfo {author} {\bibfnamefont
	  {J.}~\bibnamefont {Wang}}, \bibinfo {author} {\bibfnamefont {G.}~\bibnamefont
	  {Xue}}, \bibinfo {author} {\bibfnamefont {H.}~\bibnamefont {Yu}}, \bibinfo
	  {author} {\bibfnamefont {K.}~\bibnamefont {Huang}}, \bibinfo {author}
	  {\bibfnamefont {Z.}~\bibnamefont {Xiang}}, \bibinfo {author} {\bibfnamefont
	  {K.}~\bibnamefont {Xu}}, \bibinfo {author} {\bibfnamefont {D.}~\bibnamefont
	  {Zheng}},\ and\ \bibinfo {author} {\bibfnamefont {H.}~\bibnamefont {Fan}},\
	  }\bibfield  {title} {\bibinfo {title} {Probing spin hydrodynamics on a
	  superconducting quantum simulator},\ }\href
	  {https://doi.org/10.1038/s41467-024-52082-2} {\bibfield  {journal} {\bibinfo
	  {journal} {Nature Communications}\ }\textbf {\bibinfo {volume} {15}},\
	  \bibinfo {pages} {7573} (\bibinfo {year} {2024})}\BibitemShut {NoStop}%
	\bibitem [{\citenamefont {Sierant}\ and\ \citenamefont
	  {Zakrzewski}(2022)}]{PhysRevB.105.224203}%
	  \BibitemOpen
	  \bibfield  {author} {\bibinfo {author} {\bibfnamefont {P.}~\bibnamefont
	  {Sierant}}\ and\ \bibinfo {author} {\bibfnamefont {J.}~\bibnamefont
	  {Zakrzewski}},\ }\bibfield  {title} {\bibinfo {title} {Challenges to
	  observation of many-body localization},\ }\href
	  {https://doi.org/10.1103/PhysRevB.105.224203} {\bibfield  {journal} {\bibinfo
	   {journal} {Phys. Rev. B}\ }\textbf {\bibinfo {volume} {105}},\ \bibinfo
	  {pages} {224203} (\bibinfo {year} {2022})}\BibitemShut {NoStop}%
	\bibitem [{\citenamefont {Morningstar}\ \emph {et~al.}(2022)\citenamefont
	  {Morningstar}, \citenamefont {Colmenarez}, \citenamefont {Khemani},
	  \citenamefont {Luitz},\ and\ \citenamefont {Huse}}]{PhysRevB.105.174205}%
	  \BibitemOpen
	  \bibfield  {author} {\bibinfo {author} {\bibfnamefont {A.}~\bibnamefont
	  {Morningstar}}, \bibinfo {author} {\bibfnamefont {L.}~\bibnamefont
	  {Colmenarez}}, \bibinfo {author} {\bibfnamefont {V.}~\bibnamefont {Khemani}},
	  \bibinfo {author} {\bibfnamefont {D.~J.}\ \bibnamefont {Luitz}},\ and\
	  \bibinfo {author} {\bibfnamefont {D.~A.}\ \bibnamefont {Huse}},\ }\bibfield
	  {title} {\bibinfo {title} {Avalanches and many-body resonances in many-body
	  localized systems},\ }\href {https://doi.org/10.1103/PhysRevB.105.174205}
	  {\bibfield  {journal} {\bibinfo  {journal} {Phys. Rev. B}\ }\textbf {\bibinfo
	  {volume} {105}},\ \bibinfo {pages} {174205} (\bibinfo {year}
	  {2022})}\BibitemShut {NoStop}%
	\bibitem [{\citenamefont {{De Roeck}}\ \emph {et~al.}(2023)\citenamefont {{De
	  Roeck}}, \citenamefont {Huveneers}, \citenamefont {Meeus},\ and\
	  \citenamefont {Prośniak}}]{DEROECK2023129245}%
	  \BibitemOpen
	  \bibfield  {author} {\bibinfo {author} {\bibfnamefont {W.}~\bibnamefont {{De
	  Roeck}}}, \bibinfo {author} {\bibfnamefont {F.}~\bibnamefont {Huveneers}},
	  \bibinfo {author} {\bibfnamefont {B.}~\bibnamefont {Meeus}},\ and\ \bibinfo
	  {author} {\bibfnamefont {A.~O.}\ \bibnamefont {Prośniak}},\ }\bibfield
	  {title} {\bibinfo {title} {Rigorous and simple results on very slow
	  thermalization, or quasi-localization, of the disordered quantum chain},\
	  }\href {https://doi.org/https://doi.org/10.1016/j.physa.2023.129245}
	  {\bibfield  {journal} {\bibinfo  {journal} {Physica A: Statistical Mechanics
	  and its Applications}\ }\textbf {\bibinfo {volume} {631}},\ \bibinfo {pages}
	  {129245} (\bibinfo {year} {2023})},\ \bibinfo {note} {lecture Notes of the
	  15th International Summer School of Fundamental Problems in Statistical
	  Physics}\BibitemShut {NoStop}%
	\bibitem [{\citenamefont {De~Roeck}\ and\ \citenamefont
	  {Huveneers}(2017)}]{PhysRevB.95.155129}%
	  \BibitemOpen
	  \bibfield  {author} {\bibinfo {author} {\bibfnamefont {W.}~\bibnamefont
	  {De~Roeck}}\ and\ \bibinfo {author} {\bibfnamefont {F.~m.~c.}\ \bibnamefont
	  {Huveneers}},\ }\bibfield  {title} {\bibinfo {title} {Stability and
	  instability towards delocalization in many-body localization systems},\
	  }\href {https://doi.org/10.1103/PhysRevB.95.155129} {\bibfield  {journal}
	  {\bibinfo  {journal} {Phys. Rev. B}\ }\textbf {\bibinfo {volume} {95}},\
	  \bibinfo {pages} {155129} (\bibinfo {year} {2017})}\BibitemShut {NoStop}%
	\bibitem [{\citenamefont {Vojta}(2010)}]{Vojta2010}%
	  \BibitemOpen
	  \bibfield  {author} {\bibinfo {author} {\bibfnamefont {T.}~\bibnamefont
	  {Vojta}},\ }\bibfield  {title} {\bibinfo {title} {Quantum griffiths effects
	  and smeared phase transitions in metals: Theory and experiment},\ }\href
	  {https://doi.org/10.1007/s10909-010-0205-4} {\bibfield  {journal} {\bibinfo
	  {journal} {Journal of Low Temperature Physics}\ }\textbf {\bibinfo {volume}
	  {161}},\ \bibinfo {pages} {299} (\bibinfo {year} {2010})}\BibitemShut
	  {NoStop}%
	\bibitem [{\citenamefont {Gopalakrishnan}\ \emph {et~al.}(2016)\citenamefont
	  {Gopalakrishnan}, \citenamefont {Agarwal}, \citenamefont {Demler},
	  \citenamefont {Huse},\ and\ \citenamefont {Knap}}]{PhysRevB.93.134206}%
	  \BibitemOpen
	  \bibfield  {author} {\bibinfo {author} {\bibfnamefont {S.}~\bibnamefont
	  {Gopalakrishnan}}, \bibinfo {author} {\bibfnamefont {K.}~\bibnamefont
	  {Agarwal}}, \bibinfo {author} {\bibfnamefont {E.~A.}\ \bibnamefont {Demler}},
	  \bibinfo {author} {\bibfnamefont {D.~A.}\ \bibnamefont {Huse}},\ and\
	  \bibinfo {author} {\bibfnamefont {M.}~\bibnamefont {Knap}},\ }\bibfield
	  {title} {\bibinfo {title} {Griffiths effects and slow dynamics in nearly
	  many-body localized systems},\ }\href
	  {https://doi.org/10.1103/PhysRevB.93.134206} {\bibfield  {journal} {\bibinfo
	  {journal} {Phys. Rev. B}\ }\textbf {\bibinfo {volume} {93}},\ \bibinfo
	  {pages} {134206} (\bibinfo {year} {2016})}\BibitemShut {NoStop}%
	\bibitem [{\citenamefont {Agarwal}\ \emph {et~al.}(2015)\citenamefont
	  {Agarwal}, \citenamefont {Gopalakrishnan}, \citenamefont {Knap},
	  \citenamefont {M\"uller},\ and\ \citenamefont
	  {Demler}}]{PhysRevLett.114.160401}%
	  \BibitemOpen
	  \bibfield  {author} {\bibinfo {author} {\bibfnamefont {K.}~\bibnamefont
	  {Agarwal}}, \bibinfo {author} {\bibfnamefont {S.}~\bibnamefont
	  {Gopalakrishnan}}, \bibinfo {author} {\bibfnamefont {M.}~\bibnamefont
	  {Knap}}, \bibinfo {author} {\bibfnamefont {M.}~\bibnamefont {M\"uller}},\
	  and\ \bibinfo {author} {\bibfnamefont {E.}~\bibnamefont {Demler}},\
	  }\bibfield  {title} {\bibinfo {title} {Anomalous diffusion and griffiths
	  effects near the many-body localization transition},\ }\href
	  {https://doi.org/10.1103/PhysRevLett.114.160401} {\bibfield  {journal}
	  {\bibinfo  {journal} {Phys. Rev. Lett.}\ }\textbf {\bibinfo {volume} {114}},\
	  \bibinfo {pages} {160401} (\bibinfo {year} {2015})}\BibitemShut {NoStop}%
	\bibitem [{\citenamefont {Agarwal}\ \emph {et~al.}(2017)\citenamefont
	  {Agarwal}, \citenamefont {Altman}, \citenamefont {Demler}, \citenamefont
	  {Gopalakrishnan}, \citenamefont {Huse},\ and\ \citenamefont
	  {Knap}}]{andp.201600326}%
	  \BibitemOpen
	  \bibfield  {author} {\bibinfo {author} {\bibfnamefont {K.}~\bibnamefont
	  {Agarwal}}, \bibinfo {author} {\bibfnamefont {E.}~\bibnamefont {Altman}},
	  \bibinfo {author} {\bibfnamefont {E.}~\bibnamefont {Demler}}, \bibinfo
	  {author} {\bibfnamefont {S.}~\bibnamefont {Gopalakrishnan}}, \bibinfo
	  {author} {\bibfnamefont {D.~A.}\ \bibnamefont {Huse}},\ and\ \bibinfo
	  {author} {\bibfnamefont {M.}~\bibnamefont {Knap}},\ }\bibfield  {title}
	  {\bibinfo {title} {Rare-region effects and dynamics near the many-body
	  localization transition},\ }\href
	  {https://doi.org/https://doi.org/10.1002/andp.201600326} {\bibfield
	  {journal} {\bibinfo  {journal} {Annalen der Physik}\ }\textbf {\bibinfo
	  {volume} {529}},\ \bibinfo {pages} {1600326} (\bibinfo {year}
	  {2017})}\BibitemShut {NoStop}%
	\bibitem [{\citenamefont {Herviou}\ \emph {et~al.}(2019)\citenamefont
	  {Herviou}, \citenamefont {Bera},\ and\ \citenamefont
	  {Bardarson}}]{PhysRevB.99.134205}%
	  \BibitemOpen
	  \bibfield  {author} {\bibinfo {author} {\bibfnamefont {L.}~\bibnamefont
	  {Herviou}}, \bibinfo {author} {\bibfnamefont {S.}~\bibnamefont {Bera}},\ and\
	  \bibinfo {author} {\bibfnamefont {J.~H.}\ \bibnamefont {Bardarson}},\
	  }\bibfield  {title} {\bibinfo {title} {Multiscale entanglement clusters at
	  the many-body localization phase transition},\ }\href
	  {https://doi.org/10.1103/PhysRevB.99.134205} {\bibfield  {journal} {\bibinfo
	  {journal} {Phys. Rev. B}\ }\textbf {\bibinfo {volume} {99}},\ \bibinfo
	  {pages} {134205} (\bibinfo {year} {2019})}\BibitemShut {NoStop}%
	\bibitem [{\citenamefont {Szo\l{}dra}\ \emph {et~al.}(2021)\citenamefont
	  {Szo\l{}dra}, \citenamefont {Sierant}, \citenamefont {Kottmann},
	  \citenamefont {Lewenstein},\ and\ \citenamefont
	  {Zakrzewski}}]{PhysRevB.104.L140202}%
	  \BibitemOpen
	  \bibfield  {author} {\bibinfo {author} {\bibfnamefont {T.}~\bibnamefont
	  {Szo\l{}dra}}, \bibinfo {author} {\bibfnamefont {P.}~\bibnamefont {Sierant}},
	  \bibinfo {author} {\bibfnamefont {K.}~\bibnamefont {Kottmann}}, \bibinfo
	  {author} {\bibfnamefont {M.}~\bibnamefont {Lewenstein}},\ and\ \bibinfo
	  {author} {\bibfnamefont {J.}~\bibnamefont {Zakrzewski}},\ }\bibfield  {title}
	  {\bibinfo {title} {Detecting ergodic bubbles at the crossover to many-body
	  localization using neural networks},\ }\href
	  {https://doi.org/10.1103/PhysRevB.104.L140202} {\bibfield  {journal}
	  {\bibinfo  {journal} {Phys. Rev. B}\ }\textbf {\bibinfo {volume} {104}},\
	  \bibinfo {pages} {L140202} (\bibinfo {year} {2021})}\BibitemShut {NoStop}%
	\bibitem [{\citenamefont {Sels}(2022)}]{PhysRevB.106.L020202}%
	  \BibitemOpen
	  \bibfield  {author} {\bibinfo {author} {\bibfnamefont {D.}~\bibnamefont
	  {Sels}},\ }\bibfield  {title} {\bibinfo {title} {Bath-induced delocalization
	  in interacting disordered spin chains},\ }\href
	  {https://doi.org/10.1103/PhysRevB.106.L020202} {\bibfield  {journal}
	  {\bibinfo  {journal} {Phys. Rev. B}\ }\textbf {\bibinfo {volume} {106}},\
	  \bibinfo {pages} {L020202} (\bibinfo {year} {2022})}\BibitemShut {NoStop}%
	\bibitem [{\citenamefont {Peacock}\ and\ \citenamefont
	  {Sels}(2023)}]{PhysRevB.108.L020201}%
	  \BibitemOpen
	  \bibfield  {author} {\bibinfo {author} {\bibfnamefont {J.~C.}\ \bibnamefont
	  {Peacock}}\ and\ \bibinfo {author} {\bibfnamefont {D.}~\bibnamefont {Sels}},\
	  }\bibfield  {title} {\bibinfo {title} {Many-body delocalization from embedded
	  thermal inclusion},\ }\href {https://doi.org/10.1103/PhysRevB.108.L020201}
	  {\bibfield  {journal} {\bibinfo  {journal} {Phys. Rev. B}\ }\textbf {\bibinfo
	  {volume} {108}},\ \bibinfo {pages} {L020201} (\bibinfo {year}
	  {2023})}\BibitemShut {NoStop}%
	\bibitem [{\citenamefont {Foo}\ \emph {et~al.}(2023)\citenamefont {Foo},
	  \citenamefont {Swain}, \citenamefont {Sengupta}, \citenamefont {Lemari\'e},\
	  and\ \citenamefont {Adam}}]{PhysRevResearch.5.L032011}%
	  \BibitemOpen
	  \bibfield  {author} {\bibinfo {author} {\bibfnamefont {D.~C.~W.}\
	  \bibnamefont {Foo}}, \bibinfo {author} {\bibfnamefont {N.}~\bibnamefont
	  {Swain}}, \bibinfo {author} {\bibfnamefont {P.}~\bibnamefont {Sengupta}},
	  \bibinfo {author} {\bibfnamefont {G.}~\bibnamefont {Lemari\'e}},\ and\
	  \bibinfo {author} {\bibfnamefont {S.}~\bibnamefont {Adam}},\ }\bibfield
	  {title} {\bibinfo {title} {Stabilization mechanism for many-body localization
	  in two dimensions},\ }\href
	  {https://doi.org/10.1103/PhysRevResearch.5.L032011} {\bibfield  {journal}
	  {\bibinfo  {journal} {Phys. Rev. Res.}\ }\textbf {\bibinfo {volume} {5}},\
	  \bibinfo {pages} {L032011} (\bibinfo {year} {2023})}\BibitemShut {NoStop}%
	\bibitem [{\citenamefont {Szo\l{}dra}\ \emph {et~al.}(2024)\citenamefont
	  {Szo\l{}dra}, \citenamefont {Sierant}, \citenamefont {Lewenstein},\ and\
	  \citenamefont {Zakrzewski}}]{PhysRevB.109.134202}%
	  \BibitemOpen
	  \bibfield  {author} {\bibinfo {author} {\bibfnamefont {T.}~\bibnamefont
	  {Szo\l{}dra}}, \bibinfo {author} {\bibfnamefont {P.}~\bibnamefont {Sierant}},
	  \bibinfo {author} {\bibfnamefont {M.}~\bibnamefont {Lewenstein}},\ and\
	  \bibinfo {author} {\bibfnamefont {J.}~\bibnamefont {Zakrzewski}},\ }\bibfield
	   {title} {\bibinfo {title} {Catching thermal avalanches in the disordered xxz
	  model},\ }\href {https://doi.org/10.1103/PhysRevB.109.134202} {\bibfield
	  {journal} {\bibinfo  {journal} {Phys. Rev. B}\ }\textbf {\bibinfo {volume}
	  {109}},\ \bibinfo {pages} {134202} (\bibinfo {year} {2024})}\BibitemShut
	  {NoStop}%
	\bibitem [{\citenamefont {Kitaev}(2001)}]{AYuKitaev_2001}%
	  \BibitemOpen
	  \bibfield  {author} {\bibinfo {author} {\bibfnamefont {A.~Y.}\ \bibnamefont
	  {Kitaev}},\ }\bibfield  {title} {\bibinfo {title} {Unpaired majorana fermions
	  in quantum wires},\ }\href {https://doi.org/10.1070/1063-7869/44/10S/S29}
	  {\bibfield  {journal} {\bibinfo  {journal} {Physics-Uspekhi}\ }\textbf
	  {\bibinfo {volume} {44}},\ \bibinfo {pages} {131} (\bibinfo {year}
	  {2001})}\BibitemShut {NoStop}%
	\bibitem [{\citenamefont {Fidkowski}\ and\ \citenamefont
	  {Kitaev}(2010)}]{PhysRevB.81.134509}%
	  \BibitemOpen
	  \bibfield  {author} {\bibinfo {author} {\bibfnamefont {L.}~\bibnamefont
	  {Fidkowski}}\ and\ \bibinfo {author} {\bibfnamefont {A.}~\bibnamefont
	  {Kitaev}},\ }\bibfield  {title} {\bibinfo {title} {Effects of interactions on
	  the topological classification of free fermion systems},\ }\href
	  {https://doi.org/10.1103/PhysRevB.81.134509} {\bibfield  {journal} {\bibinfo
	  {journal} {Phys. Rev. B}\ }\textbf {\bibinfo {volume} {81}},\ \bibinfo
	  {pages} {134509} (\bibinfo {year} {2010})}\BibitemShut {NoStop}%
	\bibitem [{\citenamefont {Fidkowski}\ and\ \citenamefont
	  {Kitaev}(2011)}]{PhysRevB.83.075103}%
	  \BibitemOpen
	  \bibfield  {author} {\bibinfo {author} {\bibfnamefont {L.}~\bibnamefont
	  {Fidkowski}}\ and\ \bibinfo {author} {\bibfnamefont {A.}~\bibnamefont
	  {Kitaev}},\ }\bibfield  {title} {\bibinfo {title} {Topological phases of
	  fermions in one dimension},\ }\href
	  {https://doi.org/10.1103/PhysRevB.83.075103} {\bibfield  {journal} {\bibinfo
	  {journal} {Phys. Rev. B}\ }\textbf {\bibinfo {volume} {83}},\ \bibinfo
	  {pages} {075103} (\bibinfo {year} {2011})}\BibitemShut {NoStop}%
	\bibitem [{\citenamefont {Rahmani}\ \emph {et~al.}(2015)\citenamefont
	  {Rahmani}, \citenamefont {Zhu}, \citenamefont {Franz},\ and\ \citenamefont
	  {Affleck}}]{PhysRevLett.115.166401}%
	  \BibitemOpen
	  \bibfield  {author} {\bibinfo {author} {\bibfnamefont {A.}~\bibnamefont
	  {Rahmani}}, \bibinfo {author} {\bibfnamefont {X.}~\bibnamefont {Zhu}},
	  \bibinfo {author} {\bibfnamefont {M.}~\bibnamefont {Franz}},\ and\ \bibinfo
	  {author} {\bibfnamefont {I.}~\bibnamefont {Affleck}},\ }\bibfield  {title}
	  {\bibinfo {title} {Emergent supersymmetry from strongly interacting majorana
	  zero modes},\ }\href {https://doi.org/10.1103/PhysRevLett.115.166401}
	  {\bibfield  {journal} {\bibinfo  {journal} {Phys. Rev. Lett.}\ }\textbf
	  {\bibinfo {volume} {115}},\ \bibinfo {pages} {166401} (\bibinfo {year}
	  {2015})}\BibitemShut {NoStop}%
	\bibitem [{\citenamefont {Katsura}\ \emph {et~al.}(2015)\citenamefont
	  {Katsura}, \citenamefont {Schuricht},\ and\ \citenamefont
	  {Takahashi}}]{PhysRevB.92.115137}%
	  \BibitemOpen
	  \bibfield  {author} {\bibinfo {author} {\bibfnamefont {H.}~\bibnamefont
	  {Katsura}}, \bibinfo {author} {\bibfnamefont {D.}~\bibnamefont {Schuricht}},\
	  and\ \bibinfo {author} {\bibfnamefont {M.}~\bibnamefont {Takahashi}},\
	  }\bibfield  {title} {\bibinfo {title} {Exact ground states and topological
	  order in interacting kitaev/majorana chains},\ }\href
	  {https://doi.org/10.1103/PhysRevB.92.115137} {\bibfield  {journal} {\bibinfo
	  {journal} {Phys. Rev. B}\ }\textbf {\bibinfo {volume} {92}},\ \bibinfo
	  {pages} {115137} (\bibinfo {year} {2015})}\BibitemShut {NoStop}%
	\bibitem [{\citenamefont {Chepiga}\ and\ \citenamefont
	  {Laflorencie}(2023)}]{10.21468/SciPostPhys.14.6.152}%
	  \BibitemOpen
	  \bibfield  {author} {\bibinfo {author} {\bibfnamefont {N.}~\bibnamefont
	  {Chepiga}}\ and\ \bibinfo {author} {\bibfnamefont {N.}~\bibnamefont
	  {Laflorencie}},\ }\bibfield  {title} {\bibinfo {title} {{Topological and
	  quantum critical properties of the interacting Majorana chain model}},\
	  }\href {https://doi.org/10.21468/SciPostPhys.14.6.152} {\bibfield  {journal}
	  {\bibinfo  {journal} {SciPost Phys.}\ }\textbf {\bibinfo {volume} {14}},\
	  \bibinfo {pages} {152} (\bibinfo {year} {2023})}\BibitemShut {NoStop}%
	\bibitem [{\citenamefont {Pekker}\ \emph {et~al.}(2014)\citenamefont {Pekker},
	  \citenamefont {Refael}, \citenamefont {Altman}, \citenamefont {Demler},\ and\
	  \citenamefont {Oganesyan}}]{PhysRevX.4.011052}%
	  \BibitemOpen
	  \bibfield  {author} {\bibinfo {author} {\bibfnamefont {D.}~\bibnamefont
	  {Pekker}}, \bibinfo {author} {\bibfnamefont {G.}~\bibnamefont {Refael}},
	  \bibinfo {author} {\bibfnamefont {E.}~\bibnamefont {Altman}}, \bibinfo
	  {author} {\bibfnamefont {E.}~\bibnamefont {Demler}},\ and\ \bibinfo {author}
	  {\bibfnamefont {V.}~\bibnamefont {Oganesyan}},\ }\bibfield  {title} {\bibinfo
	  {title} {Hilbert-glass transition: New universality of temperature-tuned
	  many-body dynamical quantum criticality},\ }\href
	  {https://doi.org/10.1103/PhysRevX.4.011052} {\bibfield  {journal} {\bibinfo
	  {journal} {Phys. Rev. X}\ }\textbf {\bibinfo {volume} {4}},\ \bibinfo {pages}
	  {011052} (\bibinfo {year} {2014})}\BibitemShut {NoStop}%
	\bibitem [{\citenamefont {Kj\"all}\ \emph {et~al.}(2014)\citenamefont
	  {Kj\"all}, \citenamefont {Bardarson},\ and\ \citenamefont
	  {Pollmann}}]{PhysRevLett.113.107204}%
	  \BibitemOpen
	  \bibfield  {author} {\bibinfo {author} {\bibfnamefont {J.~A.}\ \bibnamefont
	  {Kj\"all}}, \bibinfo {author} {\bibfnamefont {J.~H.}\ \bibnamefont
	  {Bardarson}},\ and\ \bibinfo {author} {\bibfnamefont {F.}~\bibnamefont
	  {Pollmann}},\ }\bibfield  {title} {\bibinfo {title} {Many-body localization
	  in a disordered quantum ising chain},\ }\href
	  {https://doi.org/10.1103/PhysRevLett.113.107204} {\bibfield  {journal}
	  {\bibinfo  {journal} {Phys. Rev. Lett.}\ }\textbf {\bibinfo {volume} {113}},\
	  \bibinfo {pages} {107204} (\bibinfo {year} {2014})}\BibitemShut {NoStop}%
	\bibitem [{\citenamefont {Sahay}\ \emph {et~al.}(2021)\citenamefont {Sahay},
	  \citenamefont {Machado}, \citenamefont {Ye}, \citenamefont {Laumann},\ and\
	  \citenamefont {Yao}}]{PhysRevLett.126.100604}%
	  \BibitemOpen
	  \bibfield  {author} {\bibinfo {author} {\bibfnamefont {R.}~\bibnamefont
	  {Sahay}}, \bibinfo {author} {\bibfnamefont {F.}~\bibnamefont {Machado}},
	  \bibinfo {author} {\bibfnamefont {B.}~\bibnamefont {Ye}}, \bibinfo {author}
	  {\bibfnamefont {C.~R.}\ \bibnamefont {Laumann}},\ and\ \bibinfo {author}
	  {\bibfnamefont {N.~Y.}\ \bibnamefont {Yao}},\ }\bibfield  {title} {\bibinfo
	  {title} {Emergent ergodicity at the transition between many-body localized
	  phases},\ }\href {https://doi.org/10.1103/PhysRevLett.126.100604} {\bibfield
	  {journal} {\bibinfo  {journal} {Phys. Rev. Lett.}\ }\textbf {\bibinfo
	  {volume} {126}},\ \bibinfo {pages} {100604} (\bibinfo {year}
	  {2021})}\BibitemShut {NoStop}%
	\bibitem [{\citenamefont {Moudgalya}\ \emph {et~al.}(2020)\citenamefont
	  {Moudgalya}, \citenamefont {Huse},\ and\ \citenamefont
	  {Khemani}}]{moudgalya2020}%
	  \BibitemOpen
	  \bibfield  {author} {\bibinfo {author} {\bibfnamefont {S.}~\bibnamefont
	  {Moudgalya}}, \bibinfo {author} {\bibfnamefont {D.~A.}\ \bibnamefont
	  {Huse}},\ and\ \bibinfo {author} {\bibfnamefont {V.}~\bibnamefont
	  {Khemani}},\ }\href {https://arxiv.org/abs/2008.09113} {\bibinfo {title}
	  {Perturbative instability towards delocalization at phase transitions between
	  mbl phases}} (\bibinfo {year} {2020}),\ \Eprint
	  {https://arxiv.org/abs/2008.09113} {arXiv:2008.09113 [cond-mat.dis-nn]}
	  \BibitemShut {NoStop}%
	\bibitem [{\citenamefont {Laflorencie}\ \emph {et~al.}(2022)\citenamefont
	  {Laflorencie}, \citenamefont {Lemari\'e},\ and\ \citenamefont
	  {Mac\'e}}]{PhysRevResearch.4.L032016}%
	  \BibitemOpen
	  \bibfield  {author} {\bibinfo {author} {\bibfnamefont {N.}~\bibnamefont
	  {Laflorencie}}, \bibinfo {author} {\bibfnamefont {G.}~\bibnamefont
	  {Lemari\'e}},\ and\ \bibinfo {author} {\bibfnamefont {N.}~\bibnamefont
	  {Mac\'e}},\ }\bibfield  {title} {\bibinfo {title} {Topological order in
	  random interacting ising-majorana chains stabilized by many-body
	  localization},\ }\href {https://doi.org/10.1103/PhysRevResearch.4.L032016}
	  {\bibfield  {journal} {\bibinfo  {journal} {Phys. Rev. Res.}\ }\textbf
	  {\bibinfo {volume} {4}},\ \bibinfo {pages} {L032016} (\bibinfo {year}
	  {2022})}\BibitemShut {NoStop}%
	\bibitem [{\citenamefont {Thiery}\ \emph {et~al.}(2017)\citenamefont {Thiery},
	  \citenamefont {Müller},\ and\ \citenamefont
	  {Roeck}}]{thiery2017microscopicallymotivatedrenormalizationscheme}%
	  \BibitemOpen
	  \bibfield  {author} {\bibinfo {author} {\bibfnamefont {T.}~\bibnamefont
	  {Thiery}}, \bibinfo {author} {\bibfnamefont {M.}~\bibnamefont {Müller}},\
	  and\ \bibinfo {author} {\bibfnamefont {W.~D.}\ \bibnamefont {Roeck}},\ }\href
	  {https://arxiv.org/abs/1711.09880} {\bibinfo {title} {A microscopically
	  motivated renormalization scheme for the mbl/eth transition}} (\bibinfo
	  {year} {2017}),\ \Eprint {https://arxiv.org/abs/1711.09880} {arXiv:1711.09880
	  [cond-mat.stat-mech]} \BibitemShut {NoStop}%
	\bibitem [{\citenamefont {Atas}\ \emph {et~al.}(2013)\citenamefont {Atas},
	  \citenamefont {Bogomolny}, \citenamefont {Giraud},\ and\ \citenamefont
	  {Roux}}]{PhysRevLett.110.084101}%
	  \BibitemOpen
	  \bibfield  {author} {\bibinfo {author} {\bibfnamefont {Y.~Y.}\ \bibnamefont
	  {Atas}}, \bibinfo {author} {\bibfnamefont {E.}~\bibnamefont {Bogomolny}},
	  \bibinfo {author} {\bibfnamefont {O.}~\bibnamefont {Giraud}},\ and\ \bibinfo
	  {author} {\bibfnamefont {G.}~\bibnamefont {Roux}},\ }\bibfield  {title}
	  {\bibinfo {title} {Distribution of the ratio of consecutive level spacings in
	  random matrix ensembles},\ }\href
	  {https://doi.org/10.1103/PhysRevLett.110.084101} {\bibfield  {journal}
	  {\bibinfo  {journal} {Phys. Rev. Lett.}\ }\textbf {\bibinfo {volume} {110}},\
	  \bibinfo {pages} {084101} (\bibinfo {year} {2013})}\BibitemShut {NoStop}%
	\bibitem [{\citenamefont {Laflorencie}(2016)}]{LAFLORENCIE20161}%
	  \BibitemOpen
	  \bibfield  {author} {\bibinfo {author} {\bibfnamefont {N.}~\bibnamefont
	  {Laflorencie}},\ }\bibfield  {title} {\bibinfo {title} {Quantum entanglement
	  in condensed matter systems},\ }\href
	  {https://doi.org/https://doi.org/10.1016/j.physrep.2016.06.008} {\bibfield
	  {journal} {\bibinfo  {journal} {Physics Reports}\ }\textbf {\bibinfo {volume}
	  {646}},\ \bibinfo {pages} {1} (\bibinfo {year} {2016})},\ \bibinfo {note}
	  {quantum entanglement in condensed matter systems}\BibitemShut {NoStop}%
	\bibitem [{\citenamefont {Vasseur}\ \emph {et~al.}(2016)\citenamefont
	  {Vasseur}, \citenamefont {Friedman}, \citenamefont {Parameswaran},\ and\
	  \citenamefont {Potter}}]{PhysRevB.93.134207}%
	  \BibitemOpen
	  \bibfield  {author} {\bibinfo {author} {\bibfnamefont {R.}~\bibnamefont
	  {Vasseur}}, \bibinfo {author} {\bibfnamefont {A.~J.}\ \bibnamefont
	  {Friedman}}, \bibinfo {author} {\bibfnamefont {S.~A.}\ \bibnamefont
	  {Parameswaran}},\ and\ \bibinfo {author} {\bibfnamefont {A.~C.}\ \bibnamefont
	  {Potter}},\ }\bibfield  {title} {\bibinfo {title} {Particle-hole symmetry,
	  many-body localization, and topological edge modes},\ }\href
	  {https://doi.org/10.1103/PhysRevB.93.134207} {\bibfield  {journal} {\bibinfo
	  {journal} {Phys. Rev. B}\ }\textbf {\bibinfo {volume} {93}},\ \bibinfo
	  {pages} {134207} (\bibinfo {year} {2016})}\BibitemShut {NoStop}%
	\bibitem [{\citenamefont {Page}(1993)}]{PhysRevLett.71.1291}%
	  \BibitemOpen
	  \bibfield  {author} {\bibinfo {author} {\bibfnamefont {D.~N.}\ \bibnamefont
	  {Page}},\ }\bibfield  {title} {\bibinfo {title} {Average entropy of a
	  subsystem},\ }\href {https://doi.org/10.1103/PhysRevLett.71.1291} {\bibfield
	  {journal} {\bibinfo  {journal} {Phys. Rev. Lett.}\ }\textbf {\bibinfo
	  {volume} {71}},\ \bibinfo {pages} {1291} (\bibinfo {year}
	  {1993})}\BibitemShut {NoStop}%
	\bibitem [{\citenamefont {Roberts}\ and\ \citenamefont
	  {Yoshida}(2017)}]{Roberts2017}%
	  \BibitemOpen
	  \bibfield  {author} {\bibinfo {author} {\bibfnamefont {D.~A.}\ \bibnamefont
	  {Roberts}}\ and\ \bibinfo {author} {\bibfnamefont {B.}~\bibnamefont
	  {Yoshida}},\ }\bibfield  {title} {\bibinfo {title} {Chaos and complexity by
	  design},\ }\href {https://doi.org/10.1007/JHEP04(2017)121} {\bibfield
	  {journal} {\bibinfo  {journal} {Journal of High Energy Physics}\ }\textbf
	  {\bibinfo {volume} {2017}},\ \bibinfo {pages} {121} (\bibinfo {year}
	  {2017})}\BibitemShut {NoStop}%
	\bibitem [{\citenamefont {Kramers}\ and\ \citenamefont
	  {Wannier}(1941)}]{PhysRev.60.252}%
	  \BibitemOpen
	  \bibfield  {author} {\bibinfo {author} {\bibfnamefont {H.~A.}\ \bibnamefont
	  {Kramers}}\ and\ \bibinfo {author} {\bibfnamefont {G.~H.}\ \bibnamefont
	  {Wannier}},\ }\bibfield  {title} {\bibinfo {title} {Statistics of the
	  two-dimensional ferromagnet. part i},\ }\href
	  {https://doi.org/10.1103/PhysRev.60.252} {\bibfield  {journal} {\bibinfo
	  {journal} {Phys. Rev.}\ }\textbf {\bibinfo {volume} {60}},\ \bibinfo {pages}
	  {252} (\bibinfo {year} {1941})}\BibitemShut {NoStop}%
	\bibitem [{\citenamefont {Moler}\ and\ \citenamefont
	  {Van~Loan}(2003)}]{doi:10.1137/S00361445024180}%
	  \BibitemOpen
	  \bibfield  {author} {\bibinfo {author} {\bibfnamefont {C.}~\bibnamefont
	  {Moler}}\ and\ \bibinfo {author} {\bibfnamefont {C.}~\bibnamefont
	  {Van~Loan}},\ }\bibfield  {title} {\bibinfo {title} {Nineteen dubious ways to
	  compute the exponential of a matrix, twenty-five years later},\ }\href
	  {https://doi.org/10.1137/S00361445024180} {\bibfield  {journal} {\bibinfo
	  {journal} {SIAM Review}\ }\textbf {\bibinfo {volume} {45}},\ \bibinfo {pages}
	  {3} (\bibinfo {year} {2003})}\BibitemShut {NoStop}%
	\bibitem [{\citenamefont {Orell}\ \emph {et~al.}(2019)\citenamefont {Orell},
	  \citenamefont {Michailidis}, \citenamefont {Serbyn},\ and\ \citenamefont
	  {Silveri}}]{PhysRevB.100.134504}%
	  \BibitemOpen
	  \bibfield  {author} {\bibinfo {author} {\bibfnamefont {T.}~\bibnamefont
	  {Orell}}, \bibinfo {author} {\bibfnamefont {A.~A.}\ \bibnamefont
	  {Michailidis}}, \bibinfo {author} {\bibfnamefont {M.}~\bibnamefont
	  {Serbyn}},\ and\ \bibinfo {author} {\bibfnamefont {M.}~\bibnamefont
	  {Silveri}},\ }\bibfield  {title} {\bibinfo {title} {Probing the many-body
	  localization phase transition with superconducting circuits},\ }\href
	  {https://doi.org/10.1103/PhysRevB.100.134504} {\bibfield  {journal} {\bibinfo
	   {journal} {Phys. Rev. B}\ }\textbf {\bibinfo {volume} {100}},\ \bibinfo
	  {pages} {134504} (\bibinfo {year} {2019})}\BibitemShut {NoStop}%
	\end{thebibliography}

%

\end{document}